\documentclass[sigconf]{acmart}
\newcommand{\edit}[1]{\textcolor{black}{#1}}

\AtBeginDocument{
  }

\copyrightyear{2026}
\acmYear{2026}
\setcopyright{cc}
\setcctype{by}
\acmConference[CHI '26]{Proceedings of the 2026 CHI Conference on Human Factors in Computing Systems}{April 13--17, 2026}{Barcelona, Spain}
\acmBooktitle{Proceedings of the 2026 CHI Conference on Human Factors in Computing Systems (CHI '26), April 13--17, 2026, Barcelona, Spain}
\acmPrice{}
\acmDOI{10.1145/3772318.3790553}
\acmISBN{979-8-4007-2278-3/2026/04}

\usepackage{array}
\usepackage{booktabs}
\usepackage{tabularx}
\usepackage{minted}
\usepackage{subcaption}
\usepackage{hyperref}

\begin{document} 

\newcommand\todoit[1]{{\color{red}\{TODO: \textit{#1}\}}}
\newcommand\todocite{{\color{red}{CITE}}}

\definecolor{lightblue}{RGB}{212, 235, 255}
\definecolor{orange}{RGB}{255, 105, 0}
\definecolor{lightgreen}{RGB}{177, 231, 171}
\definecolor{lightyellow}{RGB}{255, 255, 148}

\newcommand\tworows[1]{\multirow{2}{*}{\shortstack[l]{#1}}}
\newcommand\tworowsc[1]{\multirow{2}{*}{\shortstack[c]{#1}}}
\newcommand\threerows[1]{\multirow{3}{*}{\shortstack[l]{#1}}}

\newcommand{\eg}{\textit{e.g.}}
\newcommand{\ie}{\textit{i.e.}}
\newcommand{\cf}{\textit{c.f.}}
\newcommand{\etal}{\textit{et al.}}
\newcommand{\system}{PaperTok}

\title[\system{}]{\system{}: Exploring the Use of Generative AI for Creating Short-form Videos for Research Communication}

\author{Meziah~Ruby~Cristobal}
\orcid{0009-0009-9473-7946}
\authornote{These authors contributed equally to this work.}
\affiliation{%
  \institution{University of Washington}
  \city{Seattle}
  \state{WA}
  \country{USA}}
\email{meziah@uw.edu}

\author{Hyeonjeong Byeon}
\orcid{0000-0002-0811-2504}
\authornotemark[1]
\affiliation{%
  \institution{University of Washington}
  \city{Seattle}
  \state{WA}
  \country{USA}}
\email{hjbyeon@uw.edu}

\author{Tze-Yu Chen}
\authornotemark[1]
\affiliation{%
  \institution{University of Washington}
  \city{Seattle}
  \state{WA}
  \country{USA}}
\email{alextyc@uw.edu}

\author{Ruoxi Shang}
\orcid{0000-0002-1062-5835}
\authornotemark[1]
\affiliation{%
  \institution{University of Washington}
  \city{Seattle}
  \state{WA}
  \country{USA}}
\email{rxshang@uw.edu}

\author{Donghoon Shin}
\orcid{0000-0001-9689-7841}
\authornotemark[1]
\affiliation{%
  \institution{University of Washington}
  \city{Seattle}
  \state{WA}
  \country{USA}}
\email{dhoon@uw.edu}

\author{Ruican Zhong}
\orcid{0009-0004-7169-0675}
\authornotemark[1]
\affiliation{%
  \institution{University of Washington}
  \city{Seattle}
  \state{WA}
  \country{USA}}
\email{rzhong98@uw.edu}

\author{Tony Zhou}
\orcid{0009-0004-2780-0419}
\authornotemark[1]
\affiliation{%
  \institution{University of Washington}
  \city{Seattle}
  \state{WA}
  \country{USA}}
\email{tyzhou05@uw.edu}

\author{Gary Hsieh}
\orcid{0000-0002-9460-2568}
\affiliation{%
  \institution{University of Washington}
  \city{Seattle}
  \state{WA}
  \country{USA}}
\email{garyhs@uw.edu}

\renewcommand{\shortauthors}{Cristobal, et al.}

\begin{abstract}
The dissemination of scholarly research is critical, yet researchers often lack the time and skills to create engaging content for popular media such as short-form videos. To address this gap, we explore the use of generative AI to help researchers transform their academic papers into accessible video content. Informed by a formative study with science communicators and content creators ($N=8$), we designed \system{}, an end-to-end system that automates the initial creative labor by generating script options and corresponding audiovisual content from a source paper. Researchers can then refine based on their preferences with further prompting. A mixed-methods user study ($N=18$) and crowdsourced evaluation ($N=100$) demonstrate that \system{}'s workflow can help researchers create engaging and informative short-form videos. We also identified the need for more fine-grained controls in the creation process. To this end, we offer implications for future generative tools that support science outreach.
\end{abstract}



\begin{teaserfigure}
  \includegraphics[width=\textwidth]{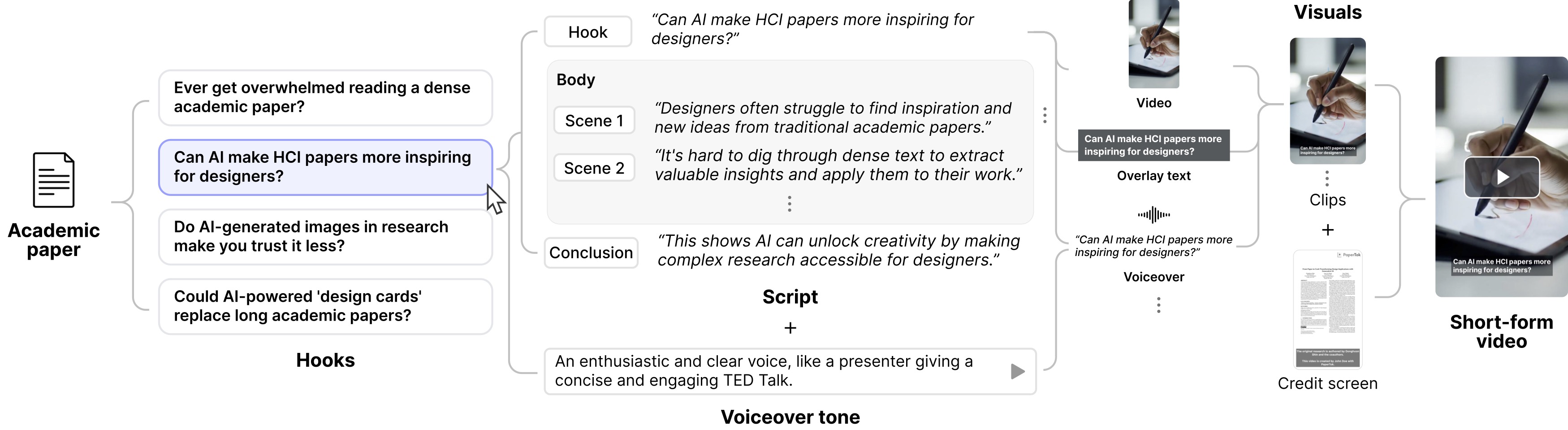}
  \caption{Overview of the \system{} system, designed to help users transform academic papers into short-form videos through an iterative workflow. Starting with a compelling hook, an accompanying script, and a suggested voiceover style, \system{} integrates AI-generated video scenes, captions, and a credit screen, streamlining the creation process.}
  \label{fig:teaser}
\end{teaserfigure}

\maketitle

\section{Introduction}


The effective communication of scholarly research is a cornerstone of scientific progress and societal well-being. \edit{However,} the primary medium for scholarly dissemination, the academic paper, is often written in \edit{a way that creates} a significant \textit{ivory tower} gap between researchers and the broader public. While there are efforts in science communication to bridge this gap, they often require significant time, resources, and a skill set in public engagement that many researchers do not possess~\cite{petzold2025bridging, williams2022hci}.

Meanwhile, the digital media landscape has fundamentally reshaped how people consume information. More than 50\% of people surveyed in a 2024 Pew report disclosed that they at least sometimes get news from social media~\cite{forman2022news}. This is further compounded by the meteoric rise of short-form video platforms (\eg{}, TikTok, Instagram Reels, YouTube Shorts). Characterized by their high engagement, visual storytelling, and algorithmic personalization, these platforms have become dominant channels for information consumption. Pew's 2025 study on TikTok usage found that 17\% of adults in the US report they regularly get news from TikTok~\cite{wang2025closer}.

Unsurprisingly, researchers and science communicators have begun to use these platforms to share scientific insights~\cite{rein2023making, wang2022assessing,frick2025science}. However, creating compelling, public-facing science content can be a formidable challenge~\cite{smith2018breaking}. There are a number of existing obstacles to communicating science~\cite{davies2008constructing, williams2022hci} that are further complicated by short-form video platforms' emphasis on brevity and highly engaging visuals. As a result, short-form video communication can impose further barriers to researchers by demanding skills in content creation (\eg{} script-writing for short-form, visualizations, narration) that might not be readily available to them.

Recent advances in generative AI have the potential to address these issues for short-form video research communication. Large language models (LLMs) can help summarize and synthesize complex information and are already being used for news reporting~\cite{shi2024generative}. Beyond text outputs, generative AI also has the potential to create audio and visual content. This can help facilitate the creation of different types of communication artifacts to support broader access and understandability of scientific content. \edit{More} recently, with text-to-video models, it is now possible to create video-based content for research communication. This new opportunity for science communication, however, raises important research questions that need to be answered. Our study explores these issues by examining: \textit{How can generative AI models be responsibly used to ensure the accuracy of the video and help improve people's understanding and trust in science? How should we design this human-AI collaborative workflow?} In addition, we also explored: \textit{What are people's perceptions of these short-form research communication videos created using generative AI?} 

While researchers have explored AI-assisted short-form video content creation~\cite{wang2024reelframer} and identified videos as a promising medium for science communication~\cite{finkler2019power}, no prior work has investigated the design and perception of AI-generated short-form videos specifically for research communication. To answer these questions, we began with a formative study with science communicators and content creators. Through a semi-structured interview using AI-generated videos as probes ($N=8$), we gained insights about how to structure these short-form videos to make them engaging and credible. Using the insights gained, we then developed \system{}, a novel system that leverages generative AI and employs a human-AI collaborative workflow to empower researchers to transform their academic papers into engaging short-form videos. Researchers upload their paper, and the system uses LLMs to analyze the paper text, identify key findings, and generate a concise, accessible video script. Subsequently, it utilizes a text-to-video model to produce a visual storyboard synchronized with the script. The system presents these AI-generated assets within an interactive editor, allowing the researcher to act as a \edit{creative} director \edit{while} ensuring scientific accuracy. This human-in-the-loop approach scaffolds the creative process, combining the efficiency of AI with the essential domain expertise and narrative voice of the human researcher. We developed \system{} to create a one-to-one translation of academic papers into short-form video to complement existing research communication practices that occur when a paper is published (\eg{}, social media \edit{posts} by researchers or university press releases). \system{} could serve as a new and engaging way to communicate science beyond traditional avenues.


We evaluated \system{} through a user study with ($N=18$) researchers who had previously published at least one academic paper. Participants were tasked with creating a short video summary of their work using \system{}. We also conducted a survey\edit{---}taken by these researcher participants and ($N=100$) crowdsourced audience participants\edit{---}to compare videos generated using \edit{\system{}} against videos generated using existing PDF-to-video platforms (\ie{}, SciSpace\footnote{https://scispace.com}, PDFtoBrainrot\footnote{https://pdftobrainrot.org}). Our findings indicate that \system{} is significantly more engaging and entertaining while providing similar levels of informational value. Researcher participants also described \system{} as a useful tool for \edit{lowering} the barrier for generating engaging visuals and voiceovers, while providing for some personalization to suit their communication style. In addition, we uncovered nuanced insights into the human-AI co-creation process, especially about researchers' desired level of control over the workflow and the need \edit{to signal the} human-in-the-loop in AI-facilitated science communication.

This paper contributes:

\begin{itemize}
\item A formative study (\S\ref{sec:formative-study}) that uncovers insights for transforming academic papers into engaging short-form videos;
\item The design and implementation of \system{} (\S\ref{sec:system-design}), a novel human-AI collaborative system for authoring short-form scientific videos from academic papers;
\item Empirical insights from a user study (\S\ref{sec:user-study}) that demonstrate the effectiveness of our system in reducing barriers to science communication and illuminate the dynamics of human-AI collaboration in a creative, expert-driven context;
\item A set of design implications (\S\ref{sec:discussion}) for future generative AI tools that aim to scaffold, rather than automate, complex creative tasks\edit{---}emphasizing the need for user control, iterative refinement, and the seamless integration of AI suggestions with human expertise.
\end{itemize}
\section{Related Work}

We outline two areas of prior work that we build on: (1) participatory web and the use of short-form videos for research communication, and (2) the potential use of generative AI for science communication. 

\subsection{Participatory Web for Research Communication}

Much research has examined the use of participatory web to communicate science~\cite{scheufele2013communicating, peters2013gap}. This includes the use of blogs~\cite{gardiner2018you, luzon2013public}, social media~\cite{darling2013role}, video-sharing sites~\cite{welbourne2016science}, podcasts~\cite{yuan2022listening}, etc. These platforms have disrupted the traditional publication pipeline~\cite{smith2018breaking,smith2020disseminating, williams2022hci} and empowered researchers to be content creators and communicate their scientific findings with the lay public directly~\cite{nisbet2009s}. For example, researchers \edit{have used} science blogs to re-contextualize research findings to be more personally relevant to readers~\cite{luzon2013public}, \edit{forums like r/science to facilitate scientific discussion forums like r/science~\cite{jones2019r}}, and \edit{social media like Twitter to foster public engagement with science}~\cite{cote2018scientists, jia2017encountered}. Not only are these efforts valuable for educating the lay public and improving society's overall science understanding, but they are also valuable to researchers themselves. Researchers communicating research on the participatory web can improve the visibility of their work and increase citation counts~\cite{lamb2018tweet}, as well as receive feedback from peers and collaborators~\cite{darling2013role}.

\edit{Video has been an important format for science communication, as combining narrative content with compelling imagery increases audience immersion and message effectiveness~\cite{finkler2019power}. As people increasingly consume} short-form videos for entertainment and information, \edit{these} platforms \edit{also become} increasingly important channels for science communication~\cite{wang2025closer}. \edit{Despite this, studies about research communication on TikTok remains} limited and \edit{typically focuses on} individual accounts~\cite{rein2023making, zawacki2022exploring}. For example, one study \edit{of 150 neuroscience TikTok videos} from @dr.brein found that shorter videos minimized viewer attrition\edit{,} and \edit{that} videos summarizing research articles \edit{outperformed} those offering advice~\cite{rein2023making}. A different study analyzed just one account again---@TerraExplore, focused on geoscience and geophysics---showed that \edit{its TikTok videos generally} received more views compared to their cross-posted videos on YouTube (Shorts) and Instagram (Reels), highlighting \edit{TikTok's algorithmic advantage for reaching audiences}~\cite{zawacki2022exploring}. 


However, while \edit{social media} can help researchers communicate \edit{broadly}, the effort required can still pose barriers for researchers. Most researchers simply do not have the time to do science outreach~\cite{williams2022hci, ecklund2012academic}. Even if they have the time, \edit{they may} lack training in translating technical jargon into plain language or in presenting their work in ways that engage diverse audiences~\cite{williams2022hci, davies2008constructing,hunter2016communications}. These challenges are now compounded when attempting to communicate through short-form videos, as researchers report lacking the time and skills to develop public-facing audios and visuals~\cite{smith2018breaking}. Combined, these challenges limit scientists' participation in short-form video science communication.   

\subsection{Rise of Generative AI and its Role in Science Communication}

Many scholars have argued that the rise of generative AI can have a transformative impact on science communication~\cite{schafer2023notorious,alvarez2024science,biyela2024generative}. \edit{Interviews with HCI scholars and other scientists have also revealed that generative AI tools are being used across research practices~\cite{kapania2025m, morris2023scientists}. Text-to-image models are used to create visuals for more engaging and understandable artifacts to communicate design implications from research findings~\cite{shin2024paper}. Text-to-audio models are used to create podcasts of scholarly papers (\eg{}, NotebookLM's Audio Overview\footnote{https://notebooklm.google/audio}). Other studies have explored the use of LLMs to generate content ideas~\cite{eldamnhory2024academify}, translate scientific jargon to laypeople~\cite{august2023paper}, and transform papers into lyrics for performable karaoke tracks~\cite{carter2025karaokai}. Similarly, prior research on text-to-video models has examined their potential for education~\cite{mittal2024comprehensive} and for patient outreach and medical training~\cite{temsah2025openai}, but not for science communication. Moreover, while short-form video is becoming increasingly influential as a medium for public engagement, its value as a science communication format---and the role of researchers in this process---has not been systematically studied.}

Commercially, there are platforms exploring the conversion of \edit{research papers in} PDF \edit{format} into videos. \edit{For example, SciSpace transforms them into video summaries, while PDFtoBrainrot converts} them into ``entertaining, TikTok-inspired \textit{brainrot} content'' to \edit{maintain attention and therefore} facilitate learning. However, neither of these tools allows for collaborative input from users during video creation. It is also unclear how people perceive these generated videos. This raises questions about the ability of these existing PDF-to-video services to support science communication, as well as how to design for researcher-AI collaboration in this context. 

Furthermore, despite the potential benefit generative AI holds for research communication, several concerns have also been noted~\cite{schafer2023notorious}. First, there are concerns about the accuracy of using generative AI to support science communication. While continuous advances in LLMs have improved on the limited quality noted in earlier models~\cite{page_why_nodate}, more current models are still susceptible to hallucination and logical fallacies~\cite{lim2024evaluation}. LLMs are also known to exhibit  biases~\cite{lin2025implicit}. These issues can be especially problematic in the context of science communication in high-stakes domains \edit{where precision matters}. Concurrently, researchers have also expressed concerns about how generative AI may contribute to the growing `infodemic' problem---where people are overloaded with excessive information~\cite{de2023chatgpt}. The lay public may experience increasing difficulty discerning between high-integrity information when overloaded with AI-generated content. \edit{These concerns highlight the need for critical inquiry into} AI’s potential role in short-form science communication \edit{and for guidelines that support accurate, trustworthy content.}

\section{Formative Study}\label{sec:formative-study}

\edit{To inform the design of \system{}, we conducted a formative study with science communication experts and content creators. Our goal was twofold: (1) document experts' current practices (\ie{}, how they plan, script, and produce short-form science communication videos), and (2) elicit their perceptions of what makes these videos engaging, understandable, and credible for target audiences. Together, these perspectives surface design requirements for tools that transform academic papers into short-form videos.}

\subsection{Participants}

We recruited 8 participants (\autoref{tab:formative-participants}) through targeted outreach to science communication content creators from Instagram, RedNote, TikTok, and YouTube, as well as university communication experts. Eligible participants were adults who create, edit, or manage original science communication for public audiences as independent creators, institutional communications staff, or researchers. We required recent original science-focused work and the ability to explain how they source and translate academic work; we excluded accounts that only reposted others’ content or were inactive channels.
\begin{table*}[h!]
\centering
\caption{Formative study participant demographics. \edit{Because P1, P4, and P7 held institutional roles managing content across multiple outlets rather than maintaining singular representative accounts, we indicated videos posted as not applicable (N/A).}}
\small
\begin{tabular}{c l l l l l}
\toprule
\textbf{PID} & \textbf{Job title} & \textbf{Main platform\edit{(s)}} & \edit{\textbf{Scicomm experience}} & \edit{\textbf{Videos posted (past year)}} \\
\midrule
P1 & Communications Manager & University and departmental news channels & \edit{12 years} & \edit{N/A} \\
P2 & Full-time Content Creator & YouTube, TikTok & \edit{5 years} & \edit{25-49} \\
P3 & Part-time Content Creator & RedNote & \edit{1 year} & \edit{100+} \\
P4 & Digital Content Creator & University news outlets & \edit{2.5 years} & \edit{N/A}\\
P5 & HCI Researcher \& Faculty & TikTok & \edit{5 years} & \edit{100+}\\
P6 & Full-time Content Creator & TikTok, YouTube, Facebook, Instagram & \edit{6 years} & \edit{100+}\\
P7 & Director of Content \& Communications & University news outlets & \edit{15 years} & \edit{N/A}\\
P8 & Part-time Content Creator & RedNote & \edit{1 year} & \edit{50-99}\\
\bottomrule
\end{tabular}
\label{tab:formative-participants}
\end{table*}

\subsection{Procedure}

We conducted 60-minute semi-structured interviews via video call, and each participant received \$30 USD in their choice of a gift card. \edit{Each session began with questions about participants' content creation workflows, challenges in translating research, and experiences with AI tools. In the second half, participants reviewed science communication video probes, evaluated their hooks, scripts, and visual concepts, and reflected on what they perceived as effective or problematic from an audience standpoint.} We concluded by co-designing improvements with participants\edit{. They rewrote} hooks, \edit{suggested} visual alternatives, and \edit{identified} missing narrative elements. Sessions were recorded, transcribed, and analyzed using thematic analysis to identify recurring patterns in quality criteria and design requirements.

\subsubsection{Video probes} \label{sec:videoprobes}
\edit{For our video probes' source papers,} we selected three CHI 2025 Best Paper awardees that each represented a different HCI research contribution ~\cite{wobbrock2016research}---artifact \cite{kim2025amuse}, empirical \cite{hsu2025placebo}, and methodological \cite{otuu2025should}. We converted \edit{each paper into a video using three methods: (1) generating a script with Google's Gemini 2.5 Flash and converting each script segment to a video clip with Veo 2 (\texttt{veo-2.0-generate-001}), then compiling them into a single short-form video; (2) using} PDFtoBrainrot, and \edit{(3) using} SciSpace\edit{'s PDF to video service. Each video was generated by the same research team member to maintain methodological consistency and enable a controlled comparison between the three generation approaches.}

PDFtoBrainrot videos consist of colorful visuals, showcasing playthroughs of popular games such as Subway Surfers and Minecraft. The scripts use slang words such as \textit{``Skibidi bop''} for humor, and captions are shown one word at a time as a bid for prolonged attention. By contrast, SciSpace primarily uses the reference paper's figures on a white background for its visuals. Its scripts are scholarly summaries, and captions are shown a few words at a time, with yellow highlights timed with the narration. Because of these stylistic differences\edit{,} we chose PDFtoBrainrot as a comparator for engagement, and SciSpace as a comparator for informativeness.

\subsection{Formative Study Findings}

In general, participants reported a desire for tools to increase their capacity. As P2 said, \textit{``the largest challenge is just time.''} Interestingly, several participants reported already using LLMs to assist with background research, and some had experimented with the use of AI-generated content in their videos as supplemental material. 

When it came to effective short-form science communication videos, participants reported several design decisions where narrative, visual, and audio elements worked together to maintain engagement and credibility. To reflect this interdependency, we present our findings as key design challenges that cut across various video elements:
\begin{itemize}
    \item Script: The narrative structure that translates academic findings into accessible language
    \begin{itemize}
        \item Hook: The opening 2--5 seconds that captures immediate attention through questions, surprising facts, or relatable pain points
    \end{itemize}
    \begin{itemize}
        \item Body: The main narrative that presents research context, key findings, etc.
    \end{itemize}
    \begin{itemize}
        \item Conclusion: The final part of the video that provides actionable takeaways and explicitly resolves the initial hook's opening question and creates narrative closure
    \end{itemize}
    \item Audio: The vocal narration that delivers the script with voiceover
    \item Visual: The imagery that illustrates the script through video clips, embedded figures, etc.
\end{itemize}

\edit{Participants’ workflow reasoning and assessments of video quality were tightly linked. For this reason, we report our findings as four cross-cutting design challenges that integrate both practices and perceptions, a structure that more directly surfaces design requirements for \system{}.}

\subsubsection{Challenge 1: Which research content should be communicated?}

Effective science communication videos must strategically select and frame research content to resonate with viewers' lived experiences while maintaining utility or educational value. This is foundational to the video generation as the initial framing stage \edit{that determines} what content should be presented.

\paragraph{Personally relevant}
The content needs to have an impact on viewers' daily lives. As P8 puts it, \textit{``How does this impact me as a person, and where am I? And like, what does this mean to me in my daily life?''} This focus on personal stakes addressed a communication challenge, as P1 observed: \textit{``People are saturated with information all the time. And so how do we get them to care about this?''} According to participants, rather than presenting abstract findings, effective videos needed to demonstrate concrete problems and solutions. P1 emphasized, \textit{``I want to show how this is a problem for people. And I want to show what our research is doing to help.''} P4 similarly prioritized identifying\edit{,} \textit{``what is the most important aspect of this in relation to like the user or just the general person.''} In addition, P1 expressed that communicating the `why'---\textit{`Why did the researcher find this to be something that they wanted to study?'}---would also be personally relevant to the audience, \textit{``because that's really going to tell people why this is something they should care about.''}

\paragraph{Timely}

Participants sought material that addressed \textit{``what's the latest''} (P1) and aligned with trending topics. P4 considered whether it is a topic  \textit{``that people are currently talking about.''} P9 selected content \textit{``close to a current hot topic, like AI's inconsistent goals in local versus national government.''} P1 recommended coordinating with news cycles: \textit{``If there's something that is kind of in the news... you should check out what our researchers are doing.''} Even specialized research needs everyday relevance. P6 was \textit{``always trying to look for hooks or things that make people understand why the things we do in space matter back here.''} 


\subsubsection{Challenge 2: How can immediate attention be captured?}

The opening seconds strongly influence whether viewers continue watching or scroll past: \textit{``The issue with short-form social media is that someone will decide in 2 seconds whether they're gonna keep watching it or scroll past it.''} (P5)

\paragraph{Hook design}

\edit{Effective hooks are crucial, as} ineffective hooks \textit{``took too long to get into what the topic was about''} (P1), burying the main point. We identified \edit{the following} principles for effective hook design:

Effective hooks acknowledge common pain points. P2 praised one of the hooks from the videos we showed---\textit{``Did you ever feel like your settings don't matter''}---noting it \textit{``speaks to me as a consumer of social media.''} Universal experiences\edit{---such as} getting interrupted while working\edit{---}also provided instant recognition\edit{, helping viewers engage with the science content that followed.} 

Hooks should generate `wait, what?' moments \edit{that require} resolution (P6). P3 uses hooks that address misconceptions: \textit{``Depression and constant travel aren't contradictory - you can be depressed in Bali.''} P2 described subverting expectations by \textit{``taking something that is used in a very traditional manner, and then putting it in a very different situation.''} Unexpected statistics also grabbed attention---after they were shown the example hook of ``98\% of Nigerians distrust police,'' P6 reacted \textit{``98, that's a really high number, and that's why I gravitate towards that.''}

Hooks must be \textit{``short and punchy, but you also don't want to lack context''} (P6). Effective creators started general before going deeper, \textit{``trying to keep it pretty general at the beginning...knowing [they're] going to have time to go more in depth later''}~(P6).

\paragraph{Visual and audio quality for immediate engagement}

The participants universally dismissed SciSpace video's primarily static screenshots of figures from the paper source. P5 stated bluntly: \textit{``Just [having] the image of the paper is, you know, anyway, no one's gonna watch that.''} P2 compared it to the other video probes, stating: \textit{``I think [dynamic visuals were] more engaging than the previous one, which, just like, showed a picture of the research paper, which I don't think is exciting for anybody.''} 

In terms of audio, artificial-sounding narration triggered instant rejection. P3 reacted to the video generated by SciSpace: \textit{``This is terrible. The AI voice immediately lowers my interest.''} P8 emphasized the \edit{need to} \textit{``humanize the content''} \edit{through narration.} The opening tone needs to establish energy and authenticity immediately, with P1 noting that just hearing an `AI voice' made them think \edit{that:} \textit{``God, no one even worked on this... like this is just a mess.''} The tone should also match the energy of short-form content\edit{. To P3, this meant} avoiding academic presentation styles that was akin to \textit{``listening to a boring academic presentation.''}

Participants also suggested using text overlays that reinforced hook questions or statements, enabling viewers to process information through multiple channels simultaneously. P9 noted that \textit{``subtitles highlighting spoken words make it engaging and less tedious,''} while P4 observed that \textit{``even just having video with text over it can be effective.''} This was particularly important given platform viewing behaviors where videos often autoplay without sound. However, P3 provided perspective on their relative impact: \textit{``Content is king. [Caption] highlighting accounts for maybe 1\% of impact. Good content makes people stay even without subtitles.''}




\subsubsection{Challenge 3: How can engagement be maintained throughout?}

While hooks capture initial attention, sustaining viewer engagement throughout the video presents distinct challenges.

\paragraph{Forming a fast-paced, complete narrative}

Supporting prior findings ~\cite{rein2023making}, participants consistently identified shorter video duration as crucial for retention. P4 explained: \textit{``15 to 30 seconds is really what people are going to pay attention to. So anything beyond that, you potentially lose interest.''} While some participants extended this to \textit{``under 60 seconds''} (P4) or \textit{``around 2 minutes long''} as an absolute maximum (P3), the consensus favored the shortest possible format that could still convey meaningful information. Beyond duration, effective videos required a clear narrative that fulfilled their opening promises. Participants emphasized that conclusions must circle back to resolve the initial framing. P2 noted: \textit{``I would almost expect, at the ending, for it to have some conclusion for me as the same person who was talked to at the beginning.''} This narrative closure needs to be actionable rather than abstract. P8 suggested adding \textit{``some sort of conclusion, (...) like, so what does this mean for them? What are the next steps? What can they do?''} Thus, the challenge is in creating a complete story---from problem to findings and practical implications---within a short duration.

\paragraph{Making complex research understandable}

Participants identified challenges in making abstract academic concepts accessible within short-form constraints. P2 articulated a fundamental difficulty in communicating research that involved subjects that are not easy to visualize, such as \textit{``something subatomic, or if you're talking about some conceptual thing.''} \edit{To} address this, visual translation emerged as a potentially useful strategy for comprehension. P9 suggested using animations, stating they \textit{``would be very helpful, especially for visualizing abstract concepts or data from the paper. It would make the content more engaging and easier to understand. For example, if there's a flow chart or a process, animating it would be much better than just showing a static image.''} Beyond visual aids, participants emphasized connecting unfamiliar concepts to \edit{familiar} experiences. P3 demonstrated this approach when discussing privacy research: \textit{``Connect it to relatable examples. Hidden cameras, hotel spying, seeing yourself on adult websites. Japanese iPhones can't turn off camera sounds because of anti-spying measures. People hate being secretly filmed.''} This strategy transformed abstract privacy concerns into immediate, emotionally resonant scenarios that viewers could instantly understand.

\paragraph{Maintaining interesting visuals}

Participants wanted visuals that showed \textit{``actual results of the study''} rather than generic stock footage (P2), though achieving this specificity proved difficult. P2 observed that while some visuals were \textit{``content related,''} they were not \textit{``really visually that interesting.''} The challenge was particularly acute for abstract research, with P9 noting: \textit{``Many papers lack a clear theme for visuals, so I have to search for images that are abstract yet relevant. This takes a lot of time.''} When screenshots of academic papers appeared on screen, participants emphasized they needed a dynamic presentation rather than a static display. P4 suggested visuals should be \textit{``zoomed in on certain parts that it was referring to, or highlighted certain, like, interesting things''} to maintain visual momentum. The overarching need was for \textit{``something visual that\edit{... [can]} be told through a video format.''}~(P6)

\subsubsection{Challenge 4: How can credibility be communicated while remaining engaging?} Participants identified credibility as crucial for science communication videos, particularly given growing skepticism about AI-generated content.

\paragraph{Adding human authenticity signals}

Human presence emerged as the primary credibility signal. Participants highlighted successful creators like Cleo Abram who \textit{``does the whole voiceover, but the first half second is always her face in every single video''} (P2). This brief personal appearance was enough to establish trust for P2\edit{:} \textit{``Even if I don't know what this topic is going to be about, I know that Cleo talks about interesting science stuff, and so I'm more willing to stick around.''} Even having a human read AI-generated content \textit{``would make a world of difference. It would create more credibility. It would make it more human''} (P5). \edit{In addition,} participants valued natural imperfections. P2 noted: \textit{``I would fumble with my words (...) [but] that just makes it feel like it's not just a robot-generated spew.''}

\paragraph{Providing academic authority markers}

Participants suggested \edit{adding explicit citation, such as authorship or institutional branding (P1)}. P5 noted \edit{their absence} in the PDFtoBrainrot video probes with dismay: \textit{``There's still no credibility markers in this, and it doesn't even cite the paper.''} \edit{Moreover, visual elements needed to reinforce} academic credibility. P9 stated: \textit{``Simple, abstract, and professional designs are preferred.''} P2 distinguished between \textit{``visuals and visuals that add to the story,''} noting that random imagery without connection to the content diminished credibility.

\paragraph{Avoiding credibility pitfalls}

Artificial visuals particularly undermined credibility by signaling low effort and a lack of care. P1 observed: \textit{``It just looks overwhelmingly AI, which maybe that's okay. But, to me, when I see that I have a sense of somebody didn't spend time on it... that \edit{[it] feels kinda, like, careless.}''} \edit{In addition,} the robotic quality of AI voices, characterized by \textit{``lack of tone variation, absence of pauses, and words that almost mashed together''} (P4), immediately signaled low production effort and reduced viewer trust. P5 predicted this sensitivity would intensify: \textit{``People are going to become even more sensitive to AI slop because there's gonna be more and more of it to the point where, like, Oh, my God, it's a human! Thank God! Let me watch this.''}
\section{The \system{} System}\label{sec:system-design}

Building \edit{on our} formative work, we designed \system{}, an interactive system that \edit{streamlines} the content creation process \edit{while} supporting \edit{user creativity and information} credibility and accessibility. In this section, we describe the user flow interacting with \system{}, as well as each feature's design considerations and implementation.

\subsection{User Flow}\label{sec:user-flow}

\edit{\system{}'s workflow} guides users from content ideation to video production. \edit{LLM is used to quickly generate content from the research paper}, while human-in-the-loop checkpoints ensure human users maintain creative control.

\begin{figure*}[t]
    \centering
    \begin{subfigure}[t]{0.33\textwidth}
        \centering
        \includegraphics[width=\linewidth]{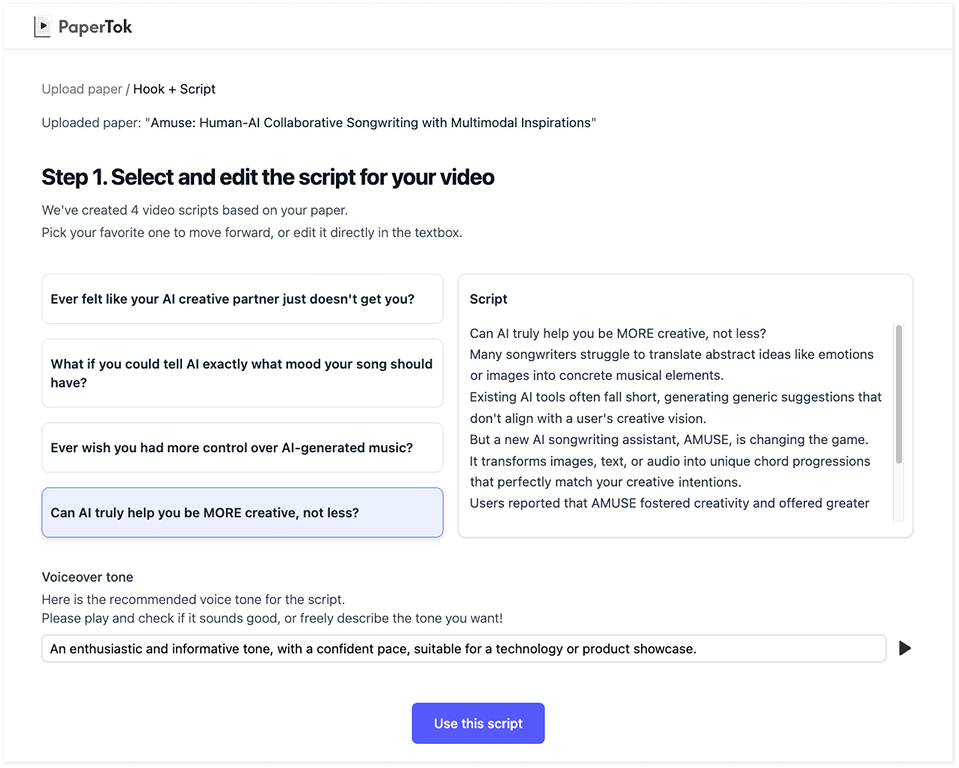}
        \caption{\textit{Hook + Script} step}
        \label{fig:script-step}
    \end{subfigure}\hfill
    \begin{subfigure}[t]{0.33\textwidth}
        \centering
        \includegraphics[width=\linewidth]{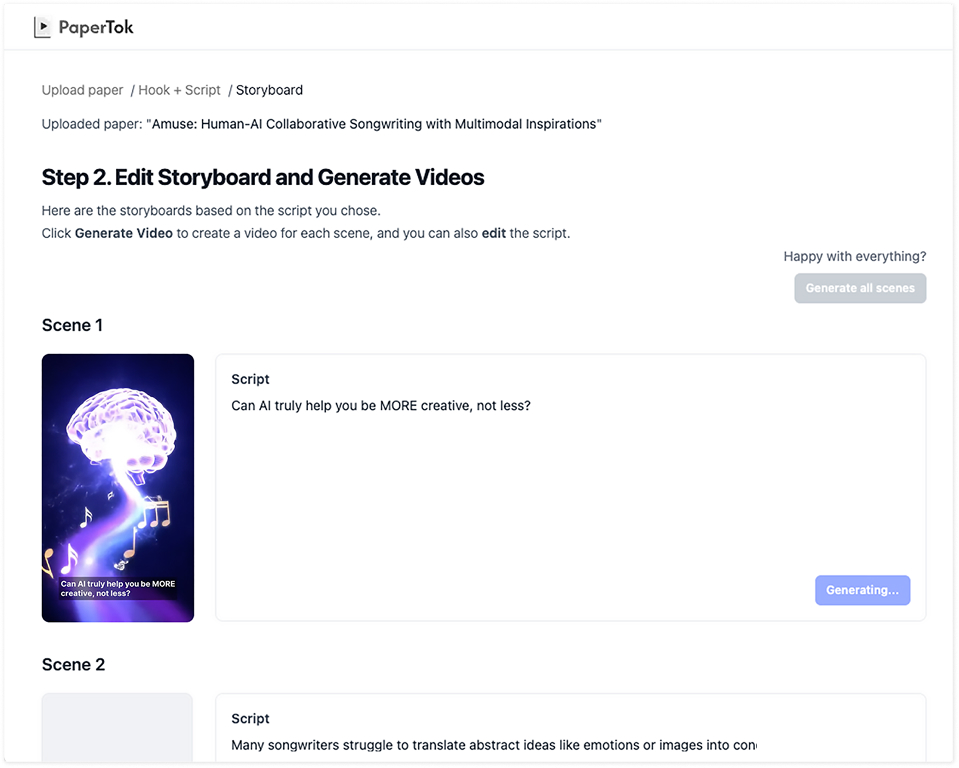}
        \caption{\textit{Storyboarding} step}
        \label{fig:storyboard-step}
    \end{subfigure}\hfill
    \begin{subfigure}[t]{0.33\textwidth}
        \centering
        \includegraphics[width=\linewidth]{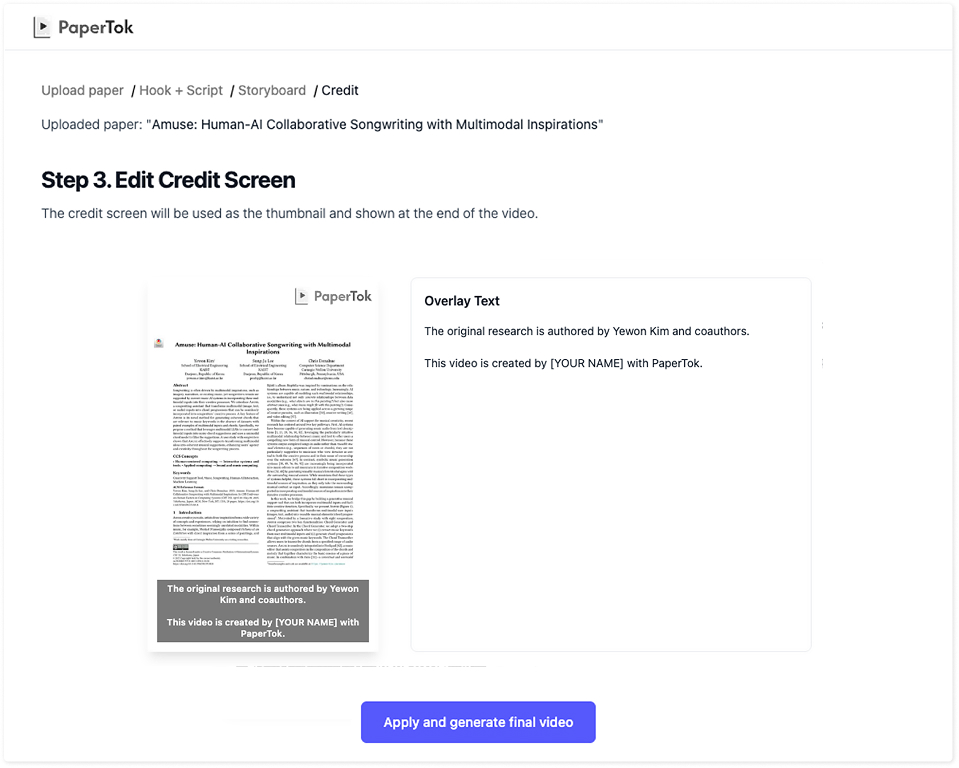}
        \caption{\textit{Credit Screen} step}
        \label{fig:credit-step}
    \end{subfigure}

    \caption{Key screens of \system{}}
    \label{fig:steps}
\end{figure*}

The process begins \edit{once} the user \edit{uploads} a PDF of their academic paper\edit{, and they are taken to \textit{Step 1: Hook + Script} (\autoref{fig:script-step}). On the left panel are four hook options with matching scripts. On the right panel is an editor where the currently selected script is displayed}, so users can \edit{refine their} narrative. \edit{Below these panels are} AI-driven recommendations for voiceover tones for each script, which users can preview and personalize by editing the prompt.

In \edit{\textit{Step 2: Storyboarding} (\autoref{fig:storyboard-step}), users are} presented with a potential script as segmented scenes, each paired with a corresponding visual to be generated. This design affords flexibility: users can generate all scenes at once then review and refine, or they can work scene by scene, refining a portion of the script and generating its visual before moving on. Users retain control throughout \edit{and engage in an iterative, storyboard-workflow, where} they can continually revise scripts and regenerate visuals until all scenes are satisfactory.

In the final step, \edit{\textit{Step 3: Credit Screen} (\autoref{fig:credit-step}),} users \edit{are prompted} to add their name in a created by-line alongside author attribution for the uploaded paper. This Credit Scene is appended to the end of the video, signaling credibility of content and the human-in-the-loop workflow.

\subsection{Script Generation}

Instead of generating a monolithic script from the start, \system{} \edit{prioritizes} the selection of a compelling hook, \edit{which guides the generation of} a script that is thematically coherent and logically flows from \edit{this} hook. In this section, we detail the design and implementation of \edit{this} process.

\subsubsection{Hook generation}

We designed a multi-step pipeline that systematically transforms academic findings into compelling hooks. This process is implemented through a system prompt that guides the LLM through a structured sequence of analysis, ideation, and refinement. The pipeline begins by instructing the LLM to identify and extract four core findings that are concrete, relatable, and have clear real-world impact. This initial step directly addresses the \edit{design requirement to} be \textit{``personally relevant''} by forcing the model to prioritize the paper's most tangible contributions over more abstract or theoretical ones\edit{. This ensures} the generated content can be engaging to audiences.

Next\edit{, the} LLM is prompted to brainstorm four distinct story ideas by applying a set of predefined creative tactics derived from our study's findings on creating cognitive tension. These tactics include leveraging contradiction (\eg{}, pitting the research against a common belief), highlighting surprise or irony, establishing personal stakes (\eg{}, \textit{`Will this replace YOUR job?'}), and creating curiosity. From a technical standpoint, this structured brainstorming prevents the LLM from defaulting to generic summaries and instead encourages it to explore multiple engaging angles for the same piece of information. For each story idea, the LLM then generates a short, compelling narrative description, which forms the basis for the final hook.

The final stage focuses on crafting punchy, clear, and engaging hooks. Here, the prompt imposes a series of strict stylistic and safety constraints. Each hook must be conversational, \edit{without} academic jargon, and limited to a maximum of 15 words to ensure brevity. A key technical consideration is the instruction to convert definitive statements into questions (\eg{}, \textit{`AI causes X'} → \textit{`Could AI cause X?'}). Through this, we aimed to mitigate the risk of oversimplifying or misrepresenting research findings, while enhancing viewer engagement by posing a question that the rest of the video promises to answer. Finally, the system includes a quality control step where the LLM is prompted to review and rate all generated hooks for engagement, relevance, and emotional appeal\edit{. Only} the top four curated options \edit{are presented} to the user. This ensures the output is not only creatively diverse but also pre-vetted for effectiveness.

\begin{table*}[h!]
\centering
\caption{\edit{Formative study findings and their implications to the design and implementation of PaperTok}}
\small
\begin{tabularx}{\textwidth}{>{\raggedright\arraybackslash}X>{\raggedright\arraybackslash}X>{\raggedright\arraybackslash}p{0.42\textwidth}}
\toprule
\textbf{Challenge} & \textbf{Design implication} & \textbf{\system{} implementation} \\
\toprule
Which research content should be communicated? & Surface claims that could be personally relevant and timely & Extracts concrete findings and generates four hook \& script options with engaging angles \\
\midrule
How can immediate attention be captured? & Include short, punchy hooks; ensure strong visuals and voice & Produces $\leq$15-word hooks framed as engaging questions; suggests relevant voiceover tones \\
\midrule
How can engagement be maintained throughout? & Create fast-paced narratives with relatable examples and visuals & Generates 8-scene scripts that follow a complete problem-solution arc, shown to the user in a storyboard-based UI \\
\midrule
How can credibility be communicated while remaining engaging? & Surface human authenticity and credibility signals & Ends video with screenshot of paper and editable, auto-filled attribution that creators can add their names to \\
\bottomrule
\end{tabularx}
\end{table*}

\subsubsection{Script}

Once the hooks are generated, \system{} proceeds to generate the full video scripts for each hook candidate. A central finding from our formative study was that a hook's effectiveness is tied to the narrative it promises; a great opening line that leads to an unrelated or confusing story results in viewer disappointment and disengagement. To address this, \system{} generates a complete \edit{script} for each of the four candidate hooks. \edit{This} process is implemented through a detailed, multi-step LLM system prompt that guides the model to produce a structured, conversational, and visually-oriented narrative\edit{.}

First, the narrative of each script is structured into eight \edit{scenes---what we found to be a good segmentation for a short-form video so that transitions are not too quick, but still fits within the typical short-form video duration of around 45 seconds}. Hook (Scene 1) introduces the premise or question. The Body (Scenes 2--7) develops the story, which consists of context (Scenes 2--3; presenting the problem or setting in a simple, relatable way), findings (Scenes 4--5; revealing the core insight, often through a visually intuitive metaphor or example), and relevance (Scenes 6--7; explaining the real-world significance, answering the audience's ``so what?''). \edit{The} Closing (Scene 8) completes the narrative loop, \edit{answering the initial} question or explaining \edit{the} surprising claim. This approach ensures a clear takeaway aligned with participant feedback. Each scene is limited to 18–22 words (6--7 seconds), supporting attention and distinct visual storytelling.

Second, we prompted the model to shape language to be both credible and accessible \edit{through} a set of rewriting rules embedded in the prompt: academic jargon is converted into everyday language, complex sentences are restructured into conversational phrasing (\eg{}, \textit{``The findings suggest...''} → \textit{``This means...''}), and abstract terms are replaced with concrete, relatable examples (\eg{}, \textit{``cellular devices''} → \textit{``your iPhone''}). To address the challenge of visualizing abstract concepts, the prompt encourages the LLM to either use tangible real-world analogies or describe concepts in ways suitable for animation. These instructions directly respond to participant feedback requesting content that feels human, expressive, and free of robotic, unemotional delivery.

Finally, the prompt embeds a self-correction step, where the LLM is instructed to review its own output against a checklist. This internal review assesses the script for coherence, conversational quality, abrupt endings, and any lingering jargon or inaccessible language. This quality assurance step ensures the final script presented to the user is polished and aligns with the platform's stylistic norms. The final output for each of the four hooks is a structured JSON array, where each object contains the scene index, script text, and a calculated duration. This machine-readable format integrates with the subsequent voice synthesis and video generation stages of the \system{} pipeline.

\subsubsection{Content-aware voice tone}

\system{} employs a customizable, two-stage voice generation process designed to balance platform-native engagement with content-specific appropriateness. Based on our formative study's findings that short-form video viewers expect a high-energy pace, we defaulted the voiceover tone as: `an influencer vibe with fast speech.' This baseline ensures that all generated audio possesses a fundamental level of energy suitable for the medium, preventing the output from sounding overly dry or academic, which participants found disengaging.

\edit{Recognizing} that a single tone cannot adequately represent the diverse emotional landscape of academic research, \edit{\system{} proactively} analyzes the script's narrative content, mood, and key message to provide a recommendation for a stylistic modifier. This is implemented in the LLM routine, which leverages an LLM to analyze the script and suggest a single, descriptive line that best complements the story (\eg{}, a paper on the detrimental effects UI dark patterns → \textit{`with an urgent, cautionary tone'}).

The user is then presented with this AI-generated recommendation within an editable text field, placing them in control of the final output. They have the agency to accept the suggestion as is, refine it to better match their vision, or discard it entirely and write a new style prompt from scratch. Once the user finalizes their stylistic preference, the LLM routine is triggered to combine the voiceover style. This client-side function takes the user's input and uses an LLM to merge it with the baseline voiceover style. The LLM is specifically instructed to preserve the core `fast speech' directive, while integrating the user's desired emotional and tonal characteristics, creating a final, consolidated prompt. For instance, if a user requests a \textit{`more serious and authoritative'} tone for a script about policy implications, \system{} converts it into \textit{``Speak fast with a serious and authoritative vibe.''} This consolidated prompt is then passed to the backend, which prepends it to the script text before sending the complete payload to the text-to-speech model.

\subsection{Video Creation}

Once the user selects a hook and finalizes the script, the workflow transitions from narrative design to visual production. This stage is meticulously designed to balance automated efficiency with fine-grained creative control, allowing users to iteratively shape the final video on a scene-by-scene basis. The process is architected as a multi-step pipeline, beginning with automated preparation, moving into an interactive user-driven generation phase, and culminating in the final assembly of the video.

\subsubsection{Script segmentation \& video prompt generation}

Before the user begins generating visuals, \system{} performs two preparatory steps in the background. First, to account for any edits, additions, or deletions made by the user in the script editor, an LLM routine re-segments the final text into discrete scene units. This ensures that the video structure remains coherent and that each segment corresponds to a manageable narrative beat, even if the user has deviated from the initial eight-scene template.

In parallel, another LLM routine generates a detailed visual prompt for each newly defined segment, converting the semantic content of the script into a set of instructions tailored for the video generation model. The prompt is generated to be highly descriptive, specifying not just the subject matter but also the desired mood, style, and camera action. For instance, \system{} translates a script line of \textit{``This means your phone is always listening''} into \textit{``A stylized, cinematic animation of abstract sound waves flowing from a person's mouth towards a glowing smartphone on a nightstand, dark and moody atmosphere, subtle dolly zoom.''}

\subsubsection{Generating each clip}

With the segments of the script and their corresponding visual prompts prepared, the user enters an interactive generation panel, which presents a storyboard view (\autoref{fig:storyboard-step}). Each segment of the script is displayed with its corresponding script text and an option to generate the visual clip. When a user initiates generation for a single scene, the frontend first sends the visual prompt, script text, and customized voice style to the backend, which immediately assigns the request a unique job identifier.

Then the backend process conducts the generation, where it first synthesizes the voiceover for the scene using the provided script and voice style. The server then calculates the duration of this resulting audio clip, which is used to dynamically set the required duration for the video clip generation. The video generation model is then invoked with the visual prompt and this exact duration, with a half-second of padding to avoid an abrupt transition to the next clip.

After the raw video is successfully generated and downloaded to the server, the system uses FFmpeg\footnote{https://ffmpeg.org/} to merge the silent video with its corresponding voiceover track. It also overlays the scene's script as burned-in text captions, a feature that improves accessibility and aligns with the established conventions of short-form video platforms.

\begin{figure*}
    \centering
    \includegraphics[width=.95\linewidth]{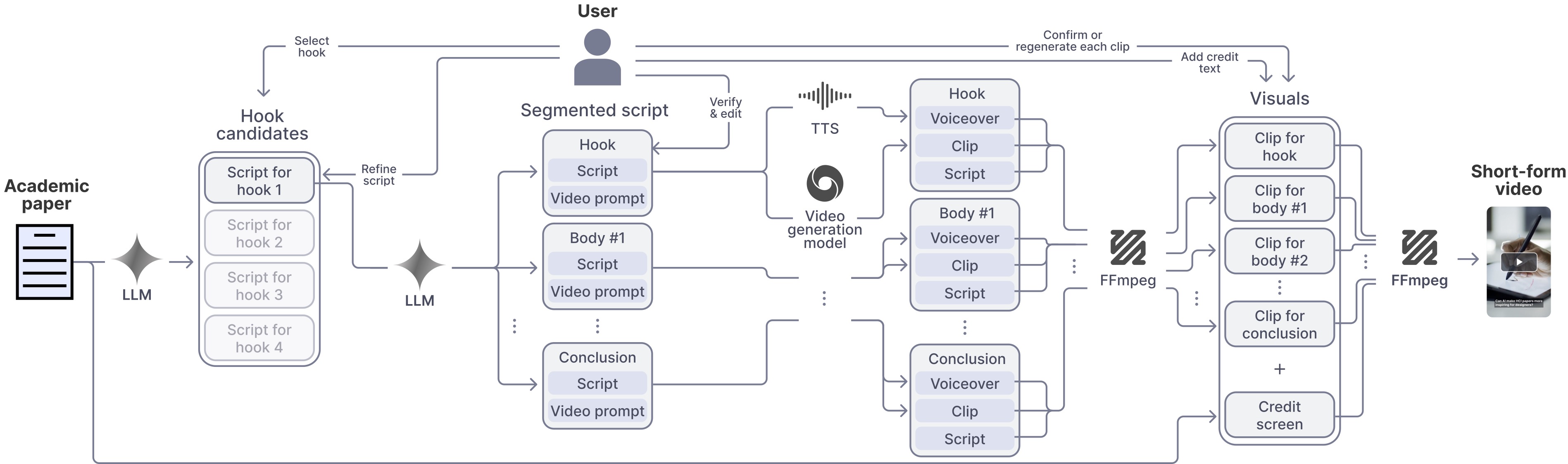}
    \caption{Overview of the \system{} system's technical pipeline and the user interaction}
    \label{fig:pipeline}
\end{figure*}

\subsection{Credit \edit{Scene as} Credibility \edit{and Human-in-the-Loop Signals}}

\edit{Drawing on formative study findings that visible human attribution enhances credibility, \system{} concludes with a} credibility-building module \edit{(\autoref{fig:credit-step})}.

\edit{To construct this ``credit scene,''} \system{} programmatically extracts a \edit{screenshot} of the paper's first page and the names of the authors. \edit{The screenshot is used as the} visual link to the original academic work, \edit{and the author names are pre-populated in the} attribution statement---\textit{``The original research is authored by \edit{<author names>}. This video is created by [Creator Name] with \system{}.''}---that is overlaid on top of the screenshot.

\edit{This attribution statement} serves the purpose of giving clear, formal credit to the original researchers, while transparently identifying the video's creator, the tool used, and the human-AI co-creation workflow. \edit{Because users} retain full flexibility to modify this text, \edit{they can also add} their own signature, social media handle, or institutional branding, \edit{allowing the final video to foreground the human creator within an otherwise AI-forward workflow}.

\subsection{Merging}

Once all individual scene clips have been generated and the credibility screen is finalized, the user triggers the final merge by clicking a button in the interface, signaling the completion of their creative process (\autoref{fig:credit-step}). This action sends a list of approved video snippets and the credibility screen data to a backend endpoint, which initiates an asynchronous merging task. The static credibility image is first converted into a two-second video clip with silent audio, and then FFmpeg's concatenation filter stitches together all scene clips and the credibility-screen video in the correct order, combining both video and audio streams into a single MP4 file. Once processing is complete, \system{} provides the user with a \edit{preview and} download link to their completed short-form video.

\subsection{Implementation}

\system{} is implemented as a web interface, where users can interact with generative models to produce video content. The interface is built on SvelteKit\footnote{https://svelte.dev/} (a JavaScript-based web app framework) and is connected to a Python backend server hosted on Google Cloud, which handles the generation and returns the results. For the generative models, we used Google's Gemini 2.5 Flash for a large language model, Veo 2 for a video generation model, and the Gemini 2.5 Flash Preview TTS model with the Zephyr prebuilt voice for a text-to-speech model. While these models were chosen as leading state-of-the-art options for their respective tasks, our system's modular design allows for the integration of alternative models into the pipeline.
\section{Study}\label{sec:user-study}
To understand the perspectives of researchers (\system{}'s target users) and the broader public (potential audiences of \system{} videos), we conducted a two-part study consisting of a survey and a user study. The survey assessed perceptions of science communication videos generated by PDF-to-video platforms, while the user study focused on evaluating researchers' experience using \system{}. Researcher participants completed both the survey and user study, while ``broader public'' participants (henceforth called ``audience'') participants only participated in the survey. We sought to address the following research questions:

\begin{enumerate}
\item[\textbf{RQ1:}] How do researchers and audiences perceive short-form science communication videos made by AI?
\item[\textbf{RQ2:}] In what ways do AI video tools support or hinder researchers' goals for communicating their work to broader audiences?
\item[\textbf{RQ3:}] What expectations and preferences do researchers have for the design of AI-enabled video tools for science communication?
\edit{\item[\textbf{RQ4:}] Are the videos generated using \system{} perceived more positively (\eg{}, credible and engaging) than those generated using available online PDF to short-form video services (\eg{}, PDFtoBrainrot, SciSpace)?}
\end{enumerate}

\subsection{Participants}

\subsubsection{Researchers.} We recruited $(N=18)$ researchers through email outreach, university Slack channels, and online scholarly communities. Prior to the study, we verified that each was a named author in at least one published research paper and that they were willing to upload one of their papers to our AI system. Participants were compensated with \$30 USD \edit{in an online gift card of their choice. These} participants are aliased as R1-18, and their roles are outlined in Table 3.

\begin{table}[h!]
\caption{\textit{Researcher} participants' roles (For more demographics information, see Appendix \ref{appendix:researcher-participants})}
\centering
\footnotesize
\begin{tabular}{c l c l}
\toprule
\textbf{PID} & \textbf{Job title} & \textbf{PID} & \textbf{Job title}\\
\midrule
R1 & Ph.D. student & R10 & Ph.D. candidate\\
R2 & Engineer, HCI researcher & R11 & Ph.D. student\\
R3 & Masters student & R12 & Ph.D. student\\
R4 & Ph.D. student & R13 & Research fellow (Ph.D. in HCI)\\
R5 & Research scientist & R14 & Ph.D. student\\
R6 & Ph.D. student & R15 & Ph.D. student\\
R7 & Ph.D. student & R16 & Postdoctoral scholar\\
R8 & Postdoctoral fellow & R17 & Ph.D. candidate\\
R9 & Ph.D. student & R18 & Ph.D. student\\
\bottomrule
\end{tabular}
\end{table}

\subsubsection{Audience.} We recruited $(N=100)$ participants using Prolific. We launched the study in 5 batches to cover different time zones and gather perspectives across the globe. Audience participants were compensated 10 USD via Prolific. \edit{These} participants are aliased as A1--A100, and their demographics information are in Appendix \ref{appendix:audience-participants}.

\subsection{Procedure} 
The following protocols were approved by our university's academic institutional review board (IRB).

\subsubsection{Survey}\label{sec:summative-survey} We designed a two-part Qualtrics survey to gather participants' feedback on\edit{: (1)} three AI-generated videos, \edit{(2)} what they find acceptable or unacceptable use of AI in science communication, and \edit{(3)} how credible and trustworthy they find AI-generated science communication videos. Similar to our formative study (see \ref{sec:videoprobes}), each participant watched a set of three videos based on the same published CHI 2025 paper, but generated by a different AI tool: PDFtoBrainrot, SciSpace, and \system{}. \edit{We chose the two comparators as they were two of the few available PDF to short-form video tools available. While both these platforms worked by simply having users upload a video, they represented two different approaches. PDFtoBrainrot, as the name suggests, overlays a voiceover summary of the paper using ``brainrot language'' on top of clips of video game play---representing a low-cost way of integrating scientific content with unrelated yet potentially engaging visuals to make the content digestible. SciSpace, on the other hand, provides a summary of the PDF with relevant images from the paper---focusing on creating video summaries of the paper. These comparators allow us to compare \system{} videos against the fun and engaging videos and the more scientific video summaries.} 

\begin{figure}
  \includegraphics[width=.9\columnwidth]{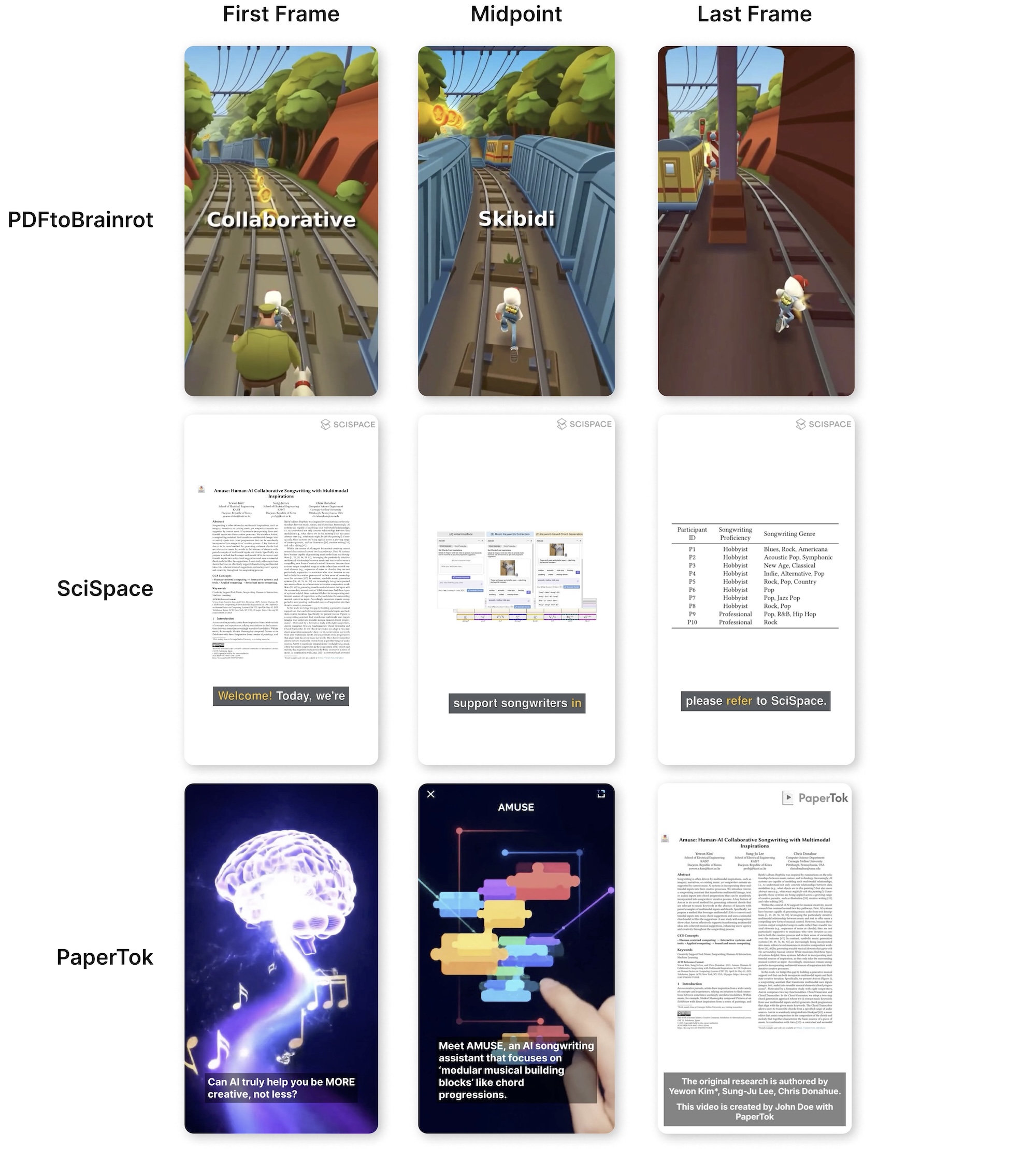}
  \caption{Screenshots from videos made using PDFtoBrainrot, SciSpace, and \system{} based on the artifact contribution paper \cite{kim2025amuse}}
  \label{fig:videos}
\end{figure}

Both participant types answered the same set of questions, except researchers selected which paper they wanted to watch videos of, while audience participants were randomly sorted into one of three sets. To prevent an ordering effect, \edit{the videos and} questions were randomized in order \edit{for} each participant.

\paragraph{Video evaluation measures} In the first part of the survey, participants evaluated each video on eleven dimensions. We measured (1) \textit{engagement}, (2) \textit{entertainment value}, and (3) \textit{informativeness} based on shared content creator and science communication success metrics we discovered in our formative study. We also measured typical social media user behaviors such as (4) \textit{shareability} and (5) \textit{intent to watch more} as additional success metrics. In addition, we measured (6) \textit{believability}, (7) \textit{accuracy}, (8) \textit{trustworthiness}, (9) \textit{bias}, and (10) \textit{completeness} to assess information credibility~\cite{flanagin2000perceptions}. Finally, we included a measure of (11) \textit{production quality} to assess overall perceptions of the video's production value.

\paragraph{AI perception measures} In the second part of the survey, we measured the trustworthiness and acceptability of AI use in science communication, based on prior work that measured similar aspects but more broadly~\cite{gerlich2023aiperceptions}. We also asked about perceptions of acceptability of AI use in specific aspects of science communication (\ie{} doing background research, creating a script, and generating visual and audio content). For our complete set of questions and measures, see Appendix \ref{appendix:survey-questions}.

\subsubsection{\system{} user study}
We conducted a 60-minute \edit{user study via video call} in pairs, with one study team member taking the lead on questions and the other taking notes on the participant's interactions with \system{}.
Each session began with the researcher participant taking the Qualtrics survey described \edit{in Section \ref{sec:summative-survey}}. During the survey, participants were provided privacy by turning off video and audio from all parties present.

\edit{After the survey}, participants shared their screen and \edit{transformed one of their authored papers into a video, following the} \system{} user flow \edit{described in Section \ref{sec:user-flow}. During lull periods when \system{} was processing (\eg{}, turning script into a segmented storyboard), the interviewer asked targeted questions to gain insight into participant choices (\eg{}, ``Why did you choose that script over the others?''). At the end of the video-making process, participants viewed the resulting product and were asked to provide reactions.}

\edit{On average, participants took about 20 minutes to generate their video. The remainder of the time was used for a} semi-structured interview about their \edit{thoughts on their \system{} video, overall} experience using \system{}\edit{,} and perceptions of AI use in science communication.

\subsection{Analyses} 

\subsubsection{Quantitative results}

To compare how participants evaluated the \edit{three different} AI-generated videos (\ie{}, PDFtoBrainrot, SciSpace, and \system{}), we built mixed-effect models for the eleven Likert scale questions, using the Likert measure as the outcome variable, and video type, paper ID, and researcher or audience as fixed effects. We modeled participant ID as the random effect due to repeated measures. For the post-hoc analyses, we report on the results from Tukey's HSD for a more conservative approach to account for the multiple pair-wise comparisons (\autoref{fig:stats}).

Additionally, we conducted a one-way ANOVA to compare audience members ($N = 100$) and researchers ($N = 18$), exploring potential differences in their perceptions of trusting an AI tool for science communication (\autoref{fig:trust}).

Lastly, to understand our audience and researcher groups' acceptability of AI usage in science communication (categorical responses), we conducted a chi-squared test of independence on audience members and researchers' responses (\autoref{fig:opinion}).

\subsubsection{Qualitative results} 
We used a hybrid approach to thematic analysis: combining a codebook~\cite{boyatzis1998codebook} with reflexive interpretation~\cite{braun2006using} to maintain methodological rigor while making space for the subjective experiences of participants and interviewers. To establish the initial codebook, four researchers independently coded the same three interviews, then met to compare interpretations and refine code definitions. After reaching consensus on the codebook, the remaining 15 interviews and 118 survey responses were divided among the four researchers for primary coding. Modifications to the codebook were encouraged and paired with asynchronous discussion in the team's Slack channel to ensure consensus while supporting the development of nuanced themes.

\section{Results}
In this section, we present the synthesized findings from the survey results and semi-structured interviews to describe how researchers and audience participants perceive AI in science communication.

\subsection{Researchers' Perceptions of \system{}} 
Creating engaging and informative science communication videos requires substantial time and effort. \system{} serves as a prototype to \edit{this challenge}, and researchers' generally positive reactions validate its intuitive interface and workflow. Furthermore, researcher reflections revealed key desirable features, like speedy workflow and clear attribution, and ways \system{} can be used to foster science communication.

\subsubsection{Functionality and value} The researcher participants' interactions with \system{} and their interview responses revealed that the system can work as a low-effort way to create science communication videos\edit{.}

\paragraph{Lowering barriers for research communication.} All 18 participants rated \system{} highly for its intuitiveness and efficient workflow. R5 remarked that anyone, \textit{``even \edit{my} mom can come in and use it without any problem.''} Participants appreciated the stepped workflow and how \textit{``with fairly minimal effort, [\system{}] was able to capture a few of the main cruxes of the paper and its contribution...for maximum information summarization''} (R1).

In addition to speed, participants also highlighted the opportunity for creating \textit{``more interesting visuals.''} (R4) This was especially helpful for R14, who found it hard to create videos themselves because they \textit{``don't have the expertise to know what visuals will be compelling... interesting to the general public.''} They further described this challenge as being difficult to figure out if they would make a visual choice that made sense to them as the author, but would \textit{``just overcomplicate it''} for \textit{``someone who is just learning about [their research topic] for the first time.''}

This difficulty in figuring out the right messaging and visuals is described by R4 as \textit{``activation energy that tools like \system{} could be helpful for.''} R4 further added: \textit{``I definitely recognize the value of having these short-form videos to disseminate content... especially to people who are probably not in academia. I definitely think that's important, but I just could not be bothered to do it myself without additional support.''}

\paragraph{Use cases} When asked what \system{} could best be used for, participants' answers revealed use cases beyond information-sharing, highlighting additional aspects of science communication tools that may be underdeveloped.
\begin{itemize}
\item{\textit{Share their research broadly. } Participants said they would use \system{} \edit{as} \textit{``an engaging way to present to a general audience''} \edit{(R16).} R13 further emphasized this point and noted it was \textit{``easy to follow... compared to presenting for 20 to 40 minutes to expert people.''} R12 believed \system{} videos could make statistical results of studies easier to understand, which could in turn make these types of research more accessible and actionable compared to now. However, they cautioned that researchers using \system{} should be \textit{``very careful''} to minimize negative impact on the audience.
}
\item{\textit{Generate ideas/support video prototyping.} Participants also mentioned that \system{} was well-suited for generating ideas for phrasing and \edit{for} visualizing their work, which they could \edit{then} take and polish in other tools they have higher degrees of control with. This was especially true for R4, R17, and R18, who had experience making science communication videos. R14 felt this inspiration could be taken beyond video, stating, \textit{``In my workflow right now, it would be most useful as kind of like an ideation thing [for] how am I going to distill my research... it might not end up in a video, [but] it might end up in a blog post, or something else.''}

R18 described \system{} as enabling rapid iteration by showcasing \textit{``diverse''} visualization styles. They also noted how helpful it would be to post multiple versions of a video and get feedback from their target audience before working on a refined final product that synthesizes the most effective elements of the video prototypes.
}

\item{\textit{Interest audience in their full paper. } Some participants felt a 45-second video is too short to fully communicate research, and instead saw it as a way to introduce broader audiences to their work and get them interested in reading the full paper. R8 recognized this challenge of getting the audience to read science, and felt that \system{} can be helpful to remedy this.
}
\item{\textit{Casually share with inner circle. } Some participants felt the current AI generation models are not \textit{``there yet''} (R5) for broader science communication. However, they envision that it could be helpful to share \system{} videos to others who are in \textit{``the general STEM society''} (R11) to tell them about their research in an engaging way.
}
\end{itemize}

\paragraph{Credit scene as a credibility signal.} A key feature of \system{} is the \edit{credit scence (\autoref{fig:videos}),} which automatically pulls the author list from the paper and generates an attribution overlay text that the user can further add to. Upon reaching this step, all 18 participants reacted positively. R1, R2, and R9 specifically mentioned this lends credibility to the video, with R9 commenting, \textit{``It shows me that it's an actual paper, not just some random stuff on Twitter.''} Researchers also liked how \textit{``the contributions are made clear.''} (R15) Some researchers suggested adding the publication venue (R11) or year (R9) to make this credit screen even better.

\subsubsection{Thoughts on AI-generated assets} 
Participants thought the models did well in developing effective hooks and engaging scripts. However, when it came to the visual elements, participants raised concerns about the text-to-video generation model's inconsistent production quality and noted this as a potential blocker for fully utilizing \system{} videos for science communication.

\paragraph{Script} \system{} provided four sets of Hook and Body options for the participant to choose from. The majority of researchers $(N=16)$ were able to use a generated script without making any modifications. When asked why they chose the script, accuracy was most commonly cited as the deciding factor. R7 even chose the script that has a \textit{``less exciting...big hook''} because \textit{``all of the information is accurate in it, which has not been totally true for the [other] scripts.''} For some researchers, this accuracy goes beyond getting facts of the paper right. For example, R5 chose a script that was also \textit{``closer to how [they] would have sold this story, how [they] would have pitched this project.''}

Half $(N=9)$ of the participants felt their research was oversimplified, but generally regarded it as a \textit{``necessary''} (R9) by-product of condensing a lengthy research paper into a short video. R11 also noted that their paper contained highly technical jargon that they do not expect those outside of the field to know, so they \textit{``wouldn't blame the video too much for [oversimplifying].''}

\paragraph{Visual.} In the Storyboard step, participants were able to generate and regenerate visuals for their scenes. The participants found most of the generated clips to be acceptable, with some even exceeding expectations in being \textit{``actually very realistic''} (R4) and depicting text correctly (R5). However, the visual style and quality proved to be inconsistent. Most participants encountered one or two visuals that had hallucinated details (\eg{} \textit{``glittery butterflies.''} in R6's clip about mixed-abilities classrooms) or were irrelevant to their script (\eg{} chains showing up in R17's clip about potential stereotypes in AI models).

More complicated, even if the individual clips were acceptable, they often differed in style and mood from each other, which made the video feel \textit{``sloppy''} (R10) once stitched together. Some participants noted that such discrepancies diminished the credibility of their work, making them unsuitable for professional presentation (R5, R11, R15).

\paragraph{Audio.} The participants understood they could modify the text-to-speech model's prompt to affect tone and delivery of the voiceover, but almost all ($N=15$) chose not to. After previewing the voiceover, participants were generally pleased and proceeded without edits. However, during the Storyboard step, some felt that the delivery and pacing of some clips could be improved, prompting R2 to wish for the same ``trial-and-error'' workflow for the audio generation, in conjunction with the visual and script iteration in the Storyboard step.

\subsection{Perceptions of AI-Generated Videos} 

To \edit{evaluate how }researchers and audiences \edit{perceived AI use in science communication and the resulting AI-generated videos,} we \edit{analyzed their} survey \edit{responses}.

\begin{figure*}
    \centering
    \includegraphics[width=\linewidth]{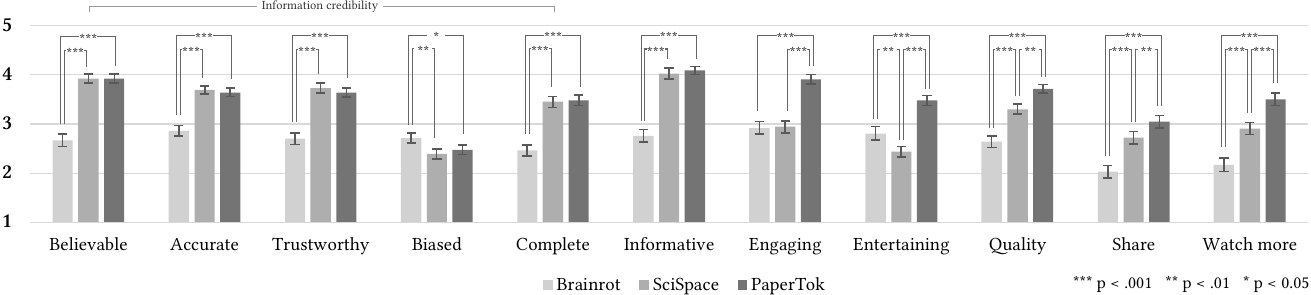}
    \caption{All participants' ($N = 118$) average ratings across 11 evaluation dimensions for each video type (PDFtoBrainrot, SciSpace, and \system{}) on our 5-point Likert scale in vertical three-column bar chart format. Significance levels from our mixed-effects model estimated means analysis are denoted above between each video type. Bars indicate standard error.
    }
    \label{fig:stats}
\end{figure*}

\system{} videos were generally positively rated and had the highest mean ratings on eight of our eleven total dimensions. Informativeness ($M = 4.09$), believability ($M = 3.92$), and engagement ($M = 3.91$) were its' three highest-rated dimensions. Willingness to share, though, \edit{while} still above neutral, had the lowest rating ($M = 3.05$, see \autoref{fig:stats}).  

\subsubsection{Video type comparisons} 

Across video platforms, we found that for the dimensions related to the informational quality of the videos, 
PDFtoBrainrot \edit{was} consistently rated the lowest compared \edit{to} SciSpace and \system{}. This included all information credibility dimensions (believable, accurate, trustworthy, biased, and complete), and the informative rating. With our post-hoc HSD tests, we showed that \system{} and SciSpace ratings were comparable (difference nonsignificant), and \edit{that} they were both significantly higher than PDFtoBrainrot's ratings along these dimensions. This suggests that videos from \system{} and SciSpace, but not PDFtoBrainrot, were positively perceived to provide informational value. 

Comparing across other dimensions more related to engagement and entertainment, we observed that \system{}'s videos were rated significantly higher compared to the other two. \system{}'s videos were rated highest in terms of engaging ($M = 3.91$), entertaining ($M = 3.48$), and quality ($M = 3.71$). Participants also reported that they were more willing to watch more ($M = 3.50$) and more likely to share ($M = 3.05$) these videos. SciSpace's ratings were comparable to PDFtoBrainrot when it came to engagement ($M = 2.94$ vs $M = 2.92, p=.98$), \edit{but} SciSpace was rated as less entertaining in comparison ($M = 2.44$ vs. $M = 2.81, p<.05$).  

\begin{figure}
    \centering\includegraphics[width=.8\linewidth]{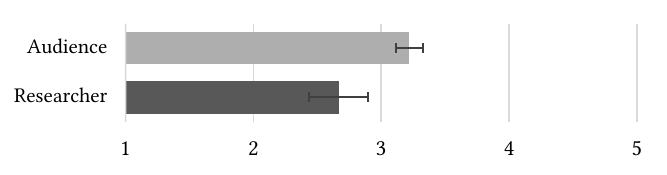}
    \caption{Audience members' and researchers' responses to \textit{``On a scale of 1 to 5, how much do you agree with this statement: I trust the use of AI for science communication.''}
    Audience members were significantly more trusting of AI use in science communication ($p < .05$) compared to researchers. Bars indicate standard error.}
    \label{fig:trust}
\end{figure}
\begin{figure}
    \centering\includegraphics[width=\linewidth]{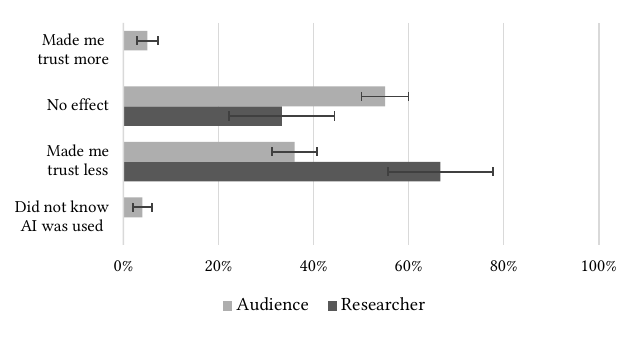}
    \caption{Audience members' and researchers' responses to \textit{``Did knowing the videos you watched were created using AI affect your opinion of them?''}
    When informed that videos were AI-generated, researchers showed a higher rate of decreased trust (66.7\%) compared to audience members (36.0\%). Audience members were more likely to respond with \textit{``No effect''} (55.0\%) compared to researchers (33.3\%). Bars indicate standard error.}
    \label{fig:opinion}
\end{figure}

\subsubsection{Audience vs. researcher perceptions}

We analyzed researcher participants' and audience participants' differences in response to our question on the level of trusting AI for science communication (Appendix \ref{appendix:survey-questions}). Our one-way ANOVA test (\autoref{fig:trust}) revealed that audience members were significantly more trustworthy of AI usage in science communication compared to researcher participants ($M = 3.22$ vs. $M = 2.67, F(1,116)=4.01, p<.05$).

We also explored whether researcher participants and audience participants differed in their evaluations of the acceptability of AI for science communication\edit{: did} knowing that the videos were AI-generated affect perceptions between the two groups\edit{?} (\autoref{fig:opinion}) Our chi-squared analysis shows that the difference between the two groups is not highly significant but trending towards significance ($\chi^{2} = 6.532$, $p < .1$). Audience participants less frequently reported 
decreased trust ($36/100$) compared to researchers ($12/18$), and \edit{were} more likely to report no effect ($55/100$ vs. $6/18$) or that it increased trust ($5/100$ \edit{vs.} none). Four audience members indicated that they did not know AI was used.

\subsection{Key Themes}
In this section, we discuss three key themes that emerge from synthesizing both the researcher and audience participants' perspectives on the role of humans in AI-facilitated science communication.

\subsubsection{Researchers desire a high level of precision in research communication} \label{sec:researchers-control}

Our findings suggest that researchers desire fine-grained control over their science communication videos. As R16 puts it: \textit{``I care a lot about this [research], so it’s important to me that the findings [in the video] are accurate.''} However, achieving this level of control is challenging due to limitations in text-to-video generation models and the nondeterministic quality of their output. This led to researchers feeling confused and frustrated when they could not determine why their script produced an undesirable visual detail. This frustration was especially difficult when they were satisfied with the script and wanted to adjust just the accompanying clip, but had no way to do it (R17). Other researchers were frustrated by common AI artifacts, such as \textit{``random, blurry text''} (R5) and \textit{``uncanny people''} (R2), that made an otherwise acceptable video unacceptable. Such \textit{``sloppy mistakes''} (R6) were considered distracting errors that undermined their work. Audience participants shared this concern, with A78 noting, \textit{``The video was good up to the point in which the vertical layout was no longer appropriate.''} \footnote{See Figure \ref{fig:vertical.png} in Appendix \ref{appendix:vid-sample} for a screenshot of the referenced clip.}

\system{} attempted to provide this \edit{level of control} by stepping users through scripting, voiceover selection, and storyboarding, but researchers wanted the option of an \textit{``expert mode''} (R2) that allowed them to prompt the AI models directly. However, researchers also recognized that prompting itself could also be a challenge: \textit{``It feels difficult to just tune the model [or] the prompt to generate something I actually expect''} \edit{(R2)}. To address existing limitations, participants proposed a range of mechanisms for increasing precision and control. Some suggested ways to shape overall style and communication strategy by providing a moodboard and customizing the voiceover audio's intonation. Others, like R13, suggested that \system{} can incorporate figures from a paper directly, like what SciSpace does in their videos, to be \textit{``more accurate in terms of information.''}

However, audience responses suggest that figures from papers should be used sparingly and carefully. While some liked the \textit{``the facts and data shown on screen to support the statements being made''} (A53), others disliked it for being \textit{``hard to follow because it just showed the research text as the video''} (A19). Even worse, A2 perceived the primarily-figures SciSpace video as \textit{``less credible, as the video is always showing image/text too small to read, making the visual part useless.''} A25 agreed, feeling that they were shown \textit{``random charts that might or might not be credible.''} From the audience's perspective, what was most critical was using these figures effectively. Even though \system{} does not currently incorporate these figures, A4 reported: \textit{``What made it more credible was the fact that it explained thoroughly what the actual topic was about, using graphics and videos that related directly.''}

\subsubsection{Researchers and audiences differ in concerns about AI videos}


Researchers often worried that audiences would dismiss their videos if they appeared \textit{``too AI-ish}'' (R2). R16 even said they would be \textit{``ashamed to share [their \system{}] video with [their] participants''} because it continued to include \textit{``nonsense text''} in the clips, despite multiple regenerations. Subtler artifacts, such as audio that sounded \textit{``unnatural, as if it didn't know what punctuation was''} (R7) or had too much \textit{``up-speak''} (R6), were also treated as threats to their credibility.

However, audience responses were more forgiving than what researchers expected. Rather than a blanket rejection, audience trust was more situational. Instead of nitpicking at AI artifacts, they judged primarily on production quality and effort. To them, the video's credibility hinged on whether the content was easy to understand (A4, A5, A10, A26, A28, A37), engaging (A7, A42, A64, A80), \edit{or} polished (A12, A57, A60, A65). In other words, audiences were not as concerned about AI use affecting the credibility of the work, unless obvious mistakes drew their attention to it (A18, A25, A97, A98).

This gap between researcher concerns and audience reactions is also reflected in their differing acceptability ratings of AI usage in the science communication process. Researchers may have a heightened sensitivity to AI use in this context because they perceive these videos as permanent and public records of their scholarship. In R6's case, the AI artifacts in their \system{} video made it \textit{``sloppy''} and \textit{``not reflective of [their] work,''}, which in turn made them reluctant to share it publicly.

These credibility concerns also extended to how authorship was represented in the credit scene. While researchers liked the attribution as paper authors, some felt uneasy being listed in a ``Created by [name] with PaperTok'' by-line. For them, ``created'' overstated their role in the process, because AI models generated the script, visuals, and audio. R5 described themselves as \textit{``more like an approver,''} and, like R7 and R13, preferred alternatives such as ``Compiled by.'' R6 even removed their name entirely---leaving only ``Created by PaperTok''---because they \textit{``don't want to take credit because the AI did it.''} Others, like R14, saw this by-line as \textit{``crucial''} context for the production quality of the video. Together, these reactions highlight how questions of authorship and ownership further compound researchers' concerns about credibility in AI-generated videos.

\subsubsection{Researchers and audience expect experts `in the loop' in AI-facilitated science communication}

Both researchers and audience participants agreed that AI could play a supportive role in science communication, but only if mechanisms are in place to ensure accuracy. Both groups acknowledged that LLMs excel in processing large amounts of information efficiently (R13, A81), but are prone to hallucination (R1, R2, R4, R7, R14, R15, A18, A25, A29, A35, A66) or \textit{``distortion''} (R3, R11, A8, A89)\edit{. Therefore,} signaling that the AI-generated video was based on rigorous scientific work and created with a human is a must. \system{} accomplished this by having a human researcher go through each step with the AI tool, in addition to the credit scene \edit{that} both participant groups appreciated as a credibility signal. However, some audience participants \edit{had higher} expectation\edit{s} and expressed that \textit{experts}\edit{---not just any human---}should be in the loop in AI-facilitated science communication (A49, A57, A98).

Researchers have mixed opinions on their responsibility in this process. While R6 felt using AI-generated videos for science communication \textit{``seems like taking away from science illustrators and human science communicators,''} R4 noted that, \textit{``AI use shouldn't be vilified,''} with the caveat that researchers are \textit{``double-checking the output.''} R17 furthered this point, stating that \textit{``AI is never reliable enough to accurately capture 100\% of the nuance needed in science communication''} so researchers must be careful to use AI only in supportive roles (R3) and to \textit{``make scientific findings''} (A20) by themselves. This shared view of using AI as an assistive tool rather than the main progenitor of ideas could be why both researchers and audiences liked the credit scene in \system{} videos and considered it a boost to the credibility (R1-18, A27, A38, A39, A46).

\edit{With regards to \textit{others} using PaperTok on their papers without their consent, participants generally viewed this as consistent with current practices---similar to journalists, bloggers, or other communicators summarizing their work. As R5 noted, \textit{``Once the research is published, it's public knowledge.''} However, as R1 and R3 requested, it would be better if \system{} could provide a notification when their work was used in PaperTok, so they can check for inaccuracies and reach out to the creator to address them.}
\section{Discussion \& Future Work}\label{sec:discussion}

Advances in generative AI create a unique opportunity to support research communication by helping generate communication artifacts that may \edit{be too difficult or costly to create otherwise}. In this work, we demonstrated that, with effective prompting and pipeline design, off-the-shelf LLMs can be leveraged to support the creation of short-form videos for research communication. Our AI-powered tool, \system{}, was able to help researchers develop an engaging script and voiceover, create visually interesting scenes, and combine everything into a shareable video with a citation. Compared to videos generated by existing PDF-to-video platforms, \system{}-created videos were rated highly for being informative \textit{and} engaging. Furthermore, although prior work has found that video-based delivery may struggle to capture different types of contributions ~\cite{williams2022hci}, we demonstrated that \system{} can generate videos for papers making artifact, empirical, and methodological contributions. 

\edit{Our findings resonate with prior work that found scientists} require precision in what is communicated~\cite{petzold2025bridging}, \edit{and that HCI researchers are} concern\edit{ed with} inaccuracies, omissions, sensationalization, and \edit{the use of} inappropriate tone \edit{for their work}~\cite{smith2020disseminating}. Importantly, our work \edit{revealed a key challenge in providing researchers with more nuanced control over AI-generated content. This} is particularly needed for \edit{video}, as, unlike text, there is no simple way to \edit{add, remove, or modify specific elements of the video without affecting the rest of it}. While direct text-to-video model prompting may help, researcher \edit{participants expressed} that fine-tuning prompts is difficult and \edit{that} current interfaces do not support such fine-grained control. This highlights \edit{a key challenge for future systems to address} in order to fully realize the potential of generative AI for research communication.

Using embedded visual content from the paper \edit{as video assets} might address this challenge. However, since these figures and images are not specifically designed for video, they could be perceived as less engaging or lacking effort, which our findings indicate can diminish audience trust. Until text-to-video models or our ability to have fine-grained control improve, future iterations of \system{}-like systems should at least provide users with the option to insert their own images or video clips that they have designed specifically for the video (\textit{e.g.}, higher resolution images, system interaction clips).

A potential concern with lowering the barrier to creating short-form research communication videos is the risk of misuse or the spread of misinformation. \edit{Our design} sought to address this in two key ways in the \edit{concluding} credit scene---a valuable credibility signal that is often missing from AI-generated content. First, we included a screenshot of the \edit{referenced academic} paper. This \edit{gives} viewers \edit{indication that the source is} peer-reviewed, while providing sufficient information to verify the source. Second, we encouraged \system{} users to ``sign-off'' on their videos by including their names and credentials. \edit{These elements reflect a broader design implication that strong credibility signals enhance the trustworthiness of the content, by making authorship and accountability visible.} In essence, our approach relies on the video creator to verify the accuracy of the information and stake their own credibility in the process.

Furthermore, this type of watermarking provides the creator acknowledgment for their work and opportunity to gain recognition for it. This branding opportunity can serve as incentive for people to create videos while ensuring the content's accuracy, as their reputation is at stake. Both researchers and audience found this to be an effective design. To further enhance trust, researcher participants suggested adding a feature in \system{} that notifies and invites authors of the papers to \edit{review} the generated video \edit{and provide feedback on accuracy and messaging}. Collectively, these \edit{design considerations point toward a model of} AI-\edit{assisted} science communication, \edit{where responsibility is distributed but traceable, and where human and AI contributions are purposefully distinguished rather than blurred.}

Finally, while our work focused on supporting the creation of short-form videos for science communication, it is important to reflect on the broader risks of AI-generated summaries of research. For example, tools like \system{} could be used to generate summaries of researchers' work without their oversight or ability to opt out, potentially leading to the dissemination of inaccurate interpretations or misrepresentations of the research. These risks are highlighted by recent changes in the ACM Digital Library, where automatic AI-generated summaries~\cite{acmdlgenai} have prompted backlash and debate over its use. Recognizing these challenges underscores the need for thoughtful design choices that balance automation with human verification and author control.

Building on this reflection, our work also points to longer-term questions that remain to be studied. While we explored the creation of videos, we did not evaluate their impact ``in-the-wild.'' Future work should explore ways to improve its reach or virality to help broaden the dissemination of these videos. \edit{Consequently, researchers should also} critically examine how this use of generative AI may transform science communication. With tools like these (and with the already exponential growth in published papers ~\cite{bornmann2021growth}), there could be a large influx of research communication content shared online. \edit{On one hand, it could replace} ``brain rot'' content---which could lead to the deterioration of a person's mental or intellectual state ~\cite{yousef2025demystifying}\edit{---and improve social media experiences. On the other hand,} short-form video content could \edit{also negatively affect and replace} people's deeper engagement with the underlying science. \edit{As short-form AI-generated research videos become more widespread, understanding how to balance information accessibility, trust, and meaningful engagement will be essential for the future of science communication.}

\section{Limitation}

While our system is adaptable to \edit{research} papers across diverse domains, our evaluation concentrated on HCI papers as a specific form of research communication. \edit{Our user study participants were also primarily Ph.D. students and early-career researchers, as they represent our target users of PaperTok.} Future work should investigate whether patterned differences in use or design implications emerge when applied to other domains \edit{or participant groups.} In addition, our system offered only text-to-speech voiceovers for audio options. While users could modify the prompt to adjust tone, other personalization factors, such as music or audio effects, were not provided for. This constrained participants' ability to fully personalize the delivery of their findings through the video, which could be a factor in lower levels of perceived ownership over the final product.



\section{Conclusion}


We introduce \system{} as an AI support tool for researchers to create accessible science communication for broader audiences. We designed a workflow that centralized \edit{the human} in key decision points, while AI handled the early, labor-intensive steps of condensing scholarly papers into a punchy, 45-second script and generating matching visuals for it. Researchers acted as creative directors as they chose, refined, and shaped their short-form video scene by scene to suit their voice and expertise.

Our studies indicate this approach \textit{can} work: science communication can be engaging for audiences and less costly for researchers. However, it also surfaced a key tension between researcher and AI when participants found value in the efficiency of the workflow but felt limited ownership when they could not precisely control visuals due to limitations in text-to-video generation. These findings suggest that the real opportunity of generative AI lies not in replacing the totality of the creative labor in creating science communication, but in enabling richer forms of human-in-the-loop collaboration with AI.

Ultimately, we see our paper contributing to the ongoing conversation about human-AI co-creation, what it means to responsibly integrate AI support tools into science communication, and how we might design systems to help people communicate and connect with research.

\begin{acks}
This project is supported by Microsoft AI and the New Future of Work Award, Google PaliGemma Academic Program GCP Credit Award, and the National Science Foundation CISE Graduate Fellowships under Grant No. G-4A-161. We would also like to thank Leah Pistorious and Minh Ton for providing invaluable feedback and support throughout this work.
\end{acks}

\bibliographystyle{ACM-Reference-Format}
\bibliography{99-bibliography}

\newpage
\onecolumn
\appendix
\section{Survey Questions}\label{appendix:survey-questions}

\subsection{Video Feedback}

The following questions in each block were randomized for each participant, except for the final question on overall production quality and final open-ended question which were fixed at the end of the survey. All questions were on a 5-point Likert scale from 1 (strongly disagree) to 5 (strongly agree), except for production quality, whose scale was from 1 (very poor quality) to 5 (very good quality).

\subsubsection{Block 1}

\begin{itemize}
    \item The video was engaging.
    \item The video was entertaining.
    \item I learned something from the video.
    \item I would share this video with others.
    \item I would watch more videos like this from the same creator.
    \item  Rate the overall production quality of the video.
\end{itemize}

\subsubsection{Block 2}

\begin{itemize}
    \item The information presented in the video was believable.
    \item The information presented in the video was accurate.
    \item The information presented in the video was trustworthy.
    \item The information presented in the video was biased.
    \item The information presented in the video was complete.
    \item What made the video seem more or less credible?
\end{itemize}

\subsection{AI Perceptions}

The following questions were based on a 5-point Likert scales were from 1 (completely disagree) to 5 (completely agree) for the trust question, and 1 (highly unacceptable) to 5 (highly acceptable) for the acceptability questions.

\begin{itemize}
    \item Did knowing the videos you watched were created using AI affect your opinion of them?
        \begin{itemize}
            \item Yes, it made me trust in the video content more
            \item Yes, it made me trust the video content less
            \item No, it didn’t affect my opinion
            \item I didn’t know AI was used until now
        \end{itemize}
    \item On a scale of 1 to 5, how much do you agree with this statement: I trust the use of AI for science communication.
    \item Why do or don’t you trust the use of AI for science communication? 
    \item  On a scale of 1 to 5, how acceptable do you find the use of AI tools in each of these aspects?
        \begin{itemize}
            \item Do background research?
            \item Create a script for a science communication video?
            \item Generate visual content (images and video) for science communication?
            \item Generate audio content (images and video) for science communication?
        \end{itemize}
\end{itemize}

\subsection{Demographics}

\begin{itemize}
    \item How often do you use social media for learning or gaining information?
    \item How often do you watch short-form videos (\textit{e.g.}, Tiktok, YouTube Shorts, Instagram Reels)?
    \item How often do you read content about science or academic research?
    \item What is your age?
    \item What is your gender?
    \item What is your highest level of education?
\end{itemize}

\newpage

\section{Summative Study Participants}

\subsection{Researcher Participant Demographics}\label{appendix:researcher-participants}

\begin{table}[h!]
\centering
\caption{Researcher participant demographics ($N = 18$)}
\footnotesize

\begin{minipage}[t]{0.33\textwidth}
\centering
\begin{tabular}{l r}
\toprule
\textbf{Category} & \textbf{N (\%)} \\
\midrule
\textit{Age} & \\
\quad 18-24 & 3 (16.7\%) \\
\quad 25-34 & 15 (83.3\%) \\
\addlinespace
\textit{Gender} & \\
\quad Female & 9 (50.0\%) \\
\quad Male & 7 (38.9\%) \\
\quad Non-binary & 1 (5.6\%) \\
\quad Prefer to self-describe & 1 (5.6\%) \\
\addlinespace
\textit{Education} & \\
\quad Bachelor's degree & 5 (27.8\%) \\
\quad Master's degree & 6 (33.3\%) \\
\quad Doctoral degree & 7 (38.9\%) \\
\bottomrule
\end{tabular}
\end{minipage}%
\hspace{0.01\textwidth}%
\begin{minipage}[t]{0.33\textwidth}
\centering
\begin{tabular}{l r}
\toprule
\textbf{Behavior} & \textbf{N (\%)} \\
\midrule
\textit{Social Media for Learning} & \\
\quad 1-2 times per week & 2 (11.1\%) \\
\quad 3-4 times per week & 1 (5.6\%) \\
\quad 5-6 times per week & 3 (16.7\%) \\
\quad Once daily & 2 (11.1\%) \\
\quad 2-5 times daily & 6 (33.3\%) \\
\quad 6-9 times daily & 1 (5.6\%) \\
\quad 10-13 times daily & 2 (11.1\%) \\
\quad Hourly or more & 1 (5.6\%) \\
\addlinespace
\textit{Short-form Video Consumption} & \\
\quad 3-4 times per week & 1 (5.6\%) \\
\quad 5-6 times per week & 1 (5.6\%) \\
\quad Once daily & 4 (22.2\%) \\
\quad 2-5 times daily & 6 (33.3\%) \\
\quad 6-9 times daily & 1 (5.6\%) \\
\quad 10-13 times daily & 4 (22.2\%) \\
\quad Hourly or more & 1 (5.6\%) \\
\addlinespace
\textit{Science Content Reading} & \\
\quad 2-3 times a week & 2 (11.1\%) \\
\quad Almost every day & 8 (44.4\%) \\
\quad Every day or more & 8 (44.4\%) \\
\bottomrule
\end{tabular}
\end{minipage}

\end{table}

\subsection{Audience Participant Demographics}\label{appendix:audience-participants}

\begin{table}[h!]
\centering
\caption{Audience participant demographics ($N = 100$)}
\footnotesize

\begin{minipage}[t]{0.35\textwidth}
\centering
\begin{tabular}{l r}
\toprule
\textbf{Category} & \textbf{N (\%)} \\
\midrule
\textit{Age} & \\
\quad 18-24 & 16 (16.0\%) \\
\quad 25-34 & 34 (34.0\%) \\
\quad 35-44 & 23 (23.0\%) \\
\quad 45-54 & 15 (15.0\%) \\
\quad 55-64 & 10 (10.0\%) \\
\quad 65 and above & 2 (2.0\%) \\
\addlinespace
\textit{Gender} & \\
\quad Female & 49 (49.0\%) \\
\quad Male & 51 (51.0\%) \\
\addlinespace
\textit{Education} & \\
\quad High school or equivalent & 9 (9.0\%) \\
\quad Some college/trade school & 20 (20.0\%) \\
\quad Bachelor's degree & 48 (48.0\%) \\
\quad Master's degree & 15 (15.0\%) \\
\quad Doctoral/Professional degree & 7 (7.0\%) \\
\quad Prefer not to say & 1 (1.0\%) \\
\bottomrule
\end{tabular}
\end{minipage}%
\hspace{0.01\textwidth}%
\begin{minipage}[t]{0.35\textwidth}
\centering
\begin{tabular}{l r}
\toprule
\textbf{Behavior} & \textbf{N (\%)} \\
\midrule
\textit{Social Media for Learning} & \\
\quad Never & 4 (4.0\%) \\
\quad 1-2 times per week & 7 (7.0\%) \\
\quad 3-4 times per week & 10 (10.0\%) \\
\quad 5-6 times per week & 8 (8.0\%) \\
\quad Once daily & 16 (16.0\%) \\
\quad 2-5 times daily & 24 (24.0\%) \\
\quad 6-9 times daily & 13 (13.0\%) \\
\quad 10-13 times daily & 8 (8.0\%) \\
\quad Hourly or more & 10 (10.0\%) \\
\addlinespace
\textit{Short-form Video Consumption} & \\
\quad Never & 1 (1.0\%) \\
\quad 1-2 times per week & 9 (9.0\%) \\
\quad 3-4 times per week & 7 (7.0\%) \\
\quad 5-6 times per week & 5 (5.0\%) \\
\quad Once daily & 7 (7.0\%) \\
\quad 2-5 times daily & 24 (24.0\%) \\
\quad 6-9 times daily & 8 (8.0\%) \\
\quad 10-13 times daily & 23 (23.0\%) \\
\quad Hourly or more & 16 (16.0\%) \\
\addlinespace
\textit{Science Content Reading} & \\
\quad Never & 2 (2.0\%) \\
\quad Less than once a month & 9 (9.0\%) \\
\quad About once a month & 5 (5.0\%) \\
\quad 2-3 times a month & 18 (18.0\%) \\
\quad About once a week & 13 (13.0\%) \\
\quad 2-3 times a week & 25 (25.0\%) \\
\quad Almost every day & 21 (21.0\%) \\
\quad Every day or more & 7 (7.0\%) \\
\bottomrule
\end{tabular}
\end{minipage}

\end{table}

\newpage

\section{\system{} Video Sample}\label{appendix:vid-sample}
\begin{figure}[h]
    \centering
    \includegraphics[width=.3\linewidth]{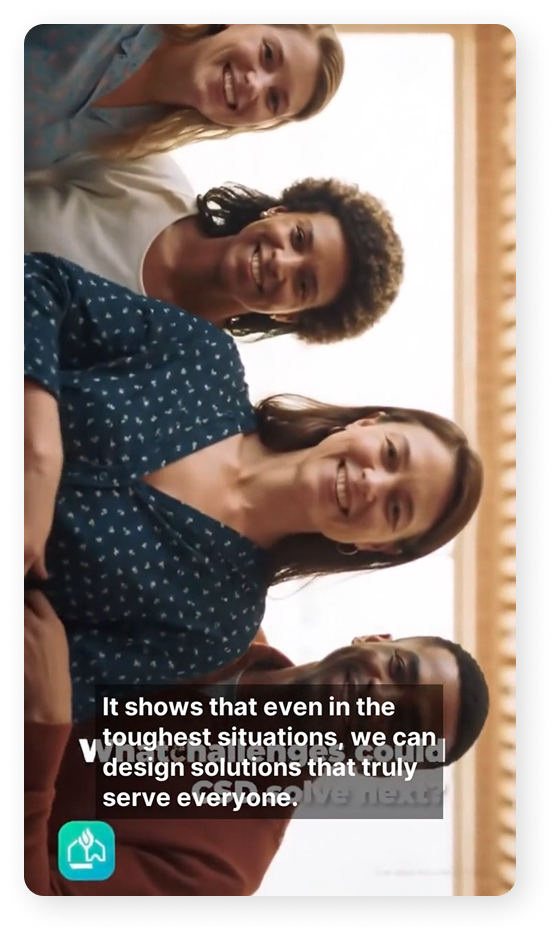}
    \caption{Screenshot of the clip referenced in \S~\ref{sec:researchers-control}}
    \label{fig:vertical.png}
\end{figure}

\end{document}